\documentclass[12pt]{article}
\usepackage{graphics}

\def\D{\displaystyle}
 \font\sevenrm=cmr7
\newif\ifinexp \inexpfalse
\newcommand\I{\ifinexp \hbox{\hskip0.8pt\sevenrm i} \else
\hbox{\hskip1pt\rm i} \fi}
\newcommand\E[1]{\inexptrue \hbox{e}^{#1} \inexpfalse}

\newcommand\re{\hbox{\hskip1pt\rm Re\hskip1pt}}
\newcommand\im{\hbox{\hskip1pt\rm Im\hskip1pt}}
\newcommand\smallre{\hbox{\hskip1pt\sevenrm Re\hskip1pt}}

\newcommand\hr{\hbox{\bf R}}
\newcommand\hc{\hbox{\bf C}}
\newcommand\diag{\,\hbox{\rm diag}}
\newcommand\tr{\,\hbox{\rm tr}}
\newcommand\Span{\,\hbox{\rm span}}
\newcommand\Image{\,\hbox{\rm Image}}

\newcommand\RRRR{{\cal R}}

\newtheorem{lemma}{\bf Lemma}
\newtheorem{theorem}{\bf Theorem}
\newtheorem{remark}{\bf Remark}
\newenvironment{demo}{\textbf{Proof. }}{\vskip4pt}

\newcommand\nar{\def\arraystretch{1}}

\def\arraystretch{1.25}
\def\fl{}
\def\dstwo{DSI\hskip-1pt I\ }
\def\dsone{DSI\ }

\title{Relation between hyperbolic Nizhnik-Novikov-Veselov equation
and stationary Davey-Stewartson I\hskip-1pt I equation}

\author{Zi-Xiang Zhou\\School of Mathematical Sciences, Fudan University,\\
Shanghai 200433, China\\
Email: zxzhou@fudan.edu.cn}

\date{}

\begin{document}

\maketitle

\begin{abstract}

A Lax system in three variables is presented, two equations of which
form the Lax pair of the stationary Davey-Stewartson I\hskip-1pt I
equation. With certain non\-linear constraints, the full
integrability condition of this Lax system contains the hyperbolic
Nizhnik-Novikov-Veselov equation and its standard Lax pair. The
Darboux transformation for the Davey-Stewartson I\hskip-1pt I
equation is used to solve the hyperbolic Nizhnik-Novikov-Veselov
equation. Using Darboux transformation, global $n$-soliton solutions
are obtained. It is proved that each $n$-soliton solution approaches
zero uniformly and exponentially at spatial infinity and is
asymptotic to $n^2$ lumps of peaks at temporal infinity.
\end{abstract}

\section{Introduction}

The Nizhnik-Novikov-Veselov (NNV) equation
\cite{bib:Nizhnik,bib:NV,bib:VN} is an important 2+1 dimensional
integrable equation which is a natural generalization of the KdV
equation to 2+1 dimensions. It is useful in both mechanics and
differential geometry \cite{bib:Kono2,bib:Kono3}. The NNV equation
has been solved by various methods such as inverse scattering
\cite{bib:Boiti}, bilinear method \cite{bib:Tagami}, bilinear
B\"acklund transformation \cite{bib:HuXB}, binary Darboux
transformation \cite{bib:Matveevbook} and so on
\cite{bib:Fordy,bib:HuLouLiu,bib:HuTangLou,bib:HuXB2,bib:Lou,bib:Nickel}.
However, one can not construct the usual Darboux transformation
(without integration) because the principal part of the first
equation of its Lax pair is two dimensional wave operator or Laplace
operator.

Starting from the idea of nonlinearization \cite{bib:Cao0}, many
high dimensional integrable systems were reduced to lower
dimensional ones so that interesting solutions like soliton
solutions and quasi-periodic solutions can be obtained from lower
dimensional systems. Especially, the KP equation
\cite{bib:CL,bib:Kono}, the \dsone equation and the $2+1$
dimensional $N$-wave equation \cite{bib:Zhou2narb} were related to
some 1+1 dimensional AKNS systems. Following this idea, in this
paper, we present a Lax system of three variables, two equations of
which form the Lax pair of the stationary
Davey-Stewartson~I\hskip-1pt I (DS I\hskip-1pt I) equation. With the
nonlinear constraints (\ref{eq:nonl_constr}), the full integrability
condition of this Lax system contains the hyperbolic NNV equation
and its standard Lax pair.

The \dstwo equation has a Darboux transformation without
integration. With the relations given by (\ref{eq:nonl_constr}), the
Darboux transformation for \dstwo equation is used to solve the
hyperbolic NNV equation. This Darboux transformation without
integration is more suitable for symbolic calculation than the known
binary Darboux transformation.

It is well known that \dsone equation has solutions approaching zero
exponentially at spatial infinity, but \dstwo equation has not.
However, we get soliton solution $u$ of the hyperbolic NNV equation
from that of the stationary \dstwo equation so that $u$ approaches
zero exponentially at spatial infinity. This is possible because the
solution $u$ of the hyperbolic NNV equation is given by $\I(g-\bar
g)$ as in (\ref{eq:nonl_constr}), not $f$, the solution of the
stationary \dstwo equation. These soliton solutions are different
from the known one derived by binary Darboux transformation or
bilinear method etc. and the behavior of the solutions is more
complicated.

In Section~\ref{sect:NNV_DS}, after reviewing the hyperbolic NNV
equation and the stationary \dstwo equation together with their
standard Lax pairs, a new Lax system (\ref{eq:LP}) is presented in
which an extra equation is added to the standard Lax pair of the
stationary \dstwo equation. With the nonlinear constraints
(\ref{eq:nonl_constr}), the integrability condition of this Lax
system includes both the hyperbolic NNV equation and its standard
Lax pair. The Darboux transformation for the new Lax system is given
in Section~\ref{sect:DT} and the general expression of multi-soliton
solutions is presented in Section~\ref{sect:stn}. In
Section~\ref{sect:1stn}, the explicit expressions and behavior of
single-soliton solutions are discussed. In Section~\ref{sect:nstn},
it is proved that each $n$-soliton solution approaches zero
uniformly and exponentially at spatial infinity. In
Section~\ref{sect:tinfty}, it is proved that each $n$-soliton
solution is asymptotic to $n^2$ lumps of peaks at temporal infinity.
Finally, some linear algebraic lemmas are presented in the Appendix.

\section{Hyperbolic Nizhnik-Novikov-Veselov equation and Davey-Stewartson I\hskip-1pt I equation}
\label{sect:NNV_DS}

The hyperbolic NNV equation is
\begin{equation}
   \begin{array}{l}
   u_t=u_{\xi\xi\xi}+u_{\eta\eta\eta}+3(uv)_{\xi}+3(uw)_{\eta},\\
   v_{\eta}=u_{\xi},\quad w_{\xi}=u_{\eta},
   \end{array}\label{eq:NNV}
\end{equation}
which has a Lax pair
\begin{equation}
   \begin{array}{l}
   \D f_{\xi\eta}+uf=0,\\
   \D f_t=f_{\xi\xi\xi}+f_{\eta\eta\eta}+3vf_{\xi}+3wf_{\eta}.
   \end{array}\label{eq:LP_NNV}
\end{equation}

By taking the new coordinates $x=\xi-\eta$, $y=\xi+\eta$, the
hyperbolic NNV equation (\ref{eq:NNV}) becomes
\begin{equation}
   \begin{array}{l}
   \D u_t=2u_{yyy}+6u_{xxy}+3(u(v+w))_y+3(u(v-w))_x,\\
   (\partial_y-\partial_x)v
   =(\partial_y+\partial_x)u,\quad (
   \partial_y+\partial_x)w
   =(\partial_y-\partial_x)u,
   \end{array}\label{eq:NNV2}
\end{equation}
and the Lax pair (\ref{eq:LP_NNV}) becomes
\begin{equation}
   \begin{array}{l}
   f_{yy}-f_{xx}+uf=0,\\
   \D f_t=2f_{yyy}+6u_{xxy}
   +3(v+w)f_y+3(v-w)f_x.
   \end{array}\label{eq:LP_NNV2}
\end{equation}

On the other hand, the \dstwo equation is
\begin{equation}
   \begin{array}{l}
   \D -\I f_\tau=f_{xx}-f_{yy}-\I(g-\bar g)f,\\
   \D(\partial_y-\I\partial_x)g=(\partial_x-\I\partial_y)(|f|^2),
   \end{array}\label{eq:DS}
\end{equation}
which has a Lax pair
\begin{equation}
   \begin{array}{l}
   \Psi_y=\I J\Psi_x+P\Psi,\\
   \D\Psi_\tau=2\I J\Psi_{xx}+2P\Psi_x+Q\Psi
    \end{array}\label{eq:LP_DS}
\end{equation}
where
\begin{equation}\fl
   \begin{array}{l}
   \D J=\left(\begin{array}{cc}1&0\\0&-1\end{array}\right),\quad
   P=\left(\begin{array}{cc}&f\\-\bar f\end{array}\right),\quad
   Q=\left(\begin{array}{cc}g &f_x-\I f_y\\
   -\bar f_x-\I\bar f_y &\bar g\end{array}\right).
   \end{array}\label{eq:JPQ}
\end{equation}

If $(f,g)$ is independent of $\tau$, (\ref{eq:DS}) becomes the
stationary \dstwo equation
\begin{equation}
   \begin{array}{l}
   \D f_{xx}-f_{yy}-\I(g-\bar g)f=0,\\
   \D(\partial_y-\I\partial_x)g=(\partial_x-\I\partial_y)(|f|^2).\\
   \end{array}\label{eq:SDS}
\end{equation}
Taking $\Psi(x,y,\tau)=\Phi(x,y)\E{2\I\lambda^2\tau}$ in
(\ref{eq:LP_DS}), we get the Lax pair for (\ref{eq:SDS}) as
\begin{equation}
   \begin{array}{l}
   \Phi_y=\I J\Phi_x+P\Phi,\\
   \D2\I\lambda^2\Phi=2\I J\Phi_{xx}+2P\Phi_x+Q\Phi
    \end{array}\label{eq:LP_SDS}
\end{equation}

The first equation of (\ref{eq:LP_NNV2}) and the first equation of
(\ref{eq:SDS}) are similar, and the second equation of
(\ref{eq:LP_NNV2}) is of order $3$. Hence we introduce an extra
equation to the Lax pair (\ref{eq:LP_SDS}) so that the whole system
becomes
\begin{equation}
   \begin{array}{l}
   \Phi_y=M(\partial)\Phi\equiv\I J\Phi_x+P\Phi,\\
   \D 2\I\lambda^2\Phi=L(\partial)\Phi\equiv 2\I J\Phi_{xx}+2P\Phi_x+Q\Phi,\\
   \Phi_t=N(\partial)\Phi\equiv 16\I J\Phi_{xxx}+16P\Phi_{xx}+R\Phi_x+S\Phi
   \end{array}\label{eq:LP}
\end{equation}
where $J$, $P$, $Q$ are given by (\ref{eq:JPQ}),
\begin{equation}\fl
   \begin{array}{l}
   \D R=4\left(\begin{array}{cc}
   3g+\I|f|^2 &4f_x-2\I f_y\\
   -4\bar f_x-2\I\bar f_y &3\bar g-\I|f|^2\end{array}\right),\\
   \D S=2\left(\begin{array}{cc}
   3g_x+2\I\bar ff_x+\bar ff_y-f\bar f_y
   &6f_{xx}-2\I f_{xy}-\I(g-\bar g)f+2|f|^2f\\
   -6\bar f_{xx}-2\I\bar f_{xy}+\I(g-\bar g)\bar f-2|f|^2\bar f
   &3\bar g_x-2\I f\bar f_x+f\bar f_y-\bar ff_y
   \end{array}\right),
   \end{array}
\end{equation}
and $L(\partial)$, $M(\partial)$ and $N(\partial)$ refer to
differential operators with respect to $x$ whose coefficients are
$2\times 2$ matrices, $\partial=\partial_x$.

The integrability conditions of (\ref{eq:LP}) include the following equations:
\begin{equation}\fl
   \begin{array}{l}
   f_{yy}-f_{xx}+uf=0,\\
   \D f_t=2f_{yyy}+6f_{xxy}+3(v+w)f_y+3(v-w)f_x,
   \end{array}\label{eq:LP_NNVb}
\end{equation}
\begin{equation}\fl
   \begin{array}{l}
   \D(\partial_y-\I\partial_x)g=(\partial_x-\I\partial_y)(|f|^2),\\
   \D\frac\I2g_t=-2g_{xxx}+2\bar ff_{xxy}+2f\bar f_{xxy}
   +4\I\bar ff_{xxx}+4\I f\bar f_{xxx}\\
   \D\qquad+2(\bar f_x-\I\bar f_y)f_{xy}
   +2(f_x-\I f_y)\bar f_{xy}
   +2(\I\bar f_x+2\bar f_y)f_{xx}
   +2(\I f_x+2f_y)\bar f_{xx}\\
   \D\qquad+(2|f|^2-\I(g-\bar g))(|f|^2)_y
   +(6\I|f|^2+(g-\bar g))(|f|^2)_x
   -2|f|^2\bar g_x+6\I gg_x,\\
   \end{array}\label{eq:SDSb}
\end{equation}
where
\begin{equation}\fl
   u=\I(g-\bar g),\quad v=2|f|^2+(g+\bar g),\quad
   w=2|f|^2-(g+\bar g).\label{eq:nonl_constr}
\end{equation}

Note that (\ref{eq:LP_NNVb}) is exactly the same as the original Lax
pair (\ref{eq:LP_NNV2}) of the hyperbolic NNV equation. By direct
calculation, we know that $(u,v,w)$ satisfies the hyperbolic NNV
equation (\ref{eq:NNV2}) provided that $f$ and $g$ satisfy
(\ref{eq:LP_NNVb})--(\ref{eq:nonl_constr}). Therefore, explicit
solutions of the hyperbolic NNV equation can be obtained from those
of (\ref{eq:LP_NNVb})--(\ref{eq:nonl_constr}).

Clearly, the solutions of (\ref{eq:LP_NNVb})--(\ref{eq:nonl_constr})
are only part of those of the hyperbolic NNV equation. However, they
include some interesting ones which will be shown in the rest of
this paper.

\section{Darboux transformation}\label{sect:DT}

The binary Darboux transformation for the hyperbolic NNV equation is
well-known \cite{bib:Matveevbook}. Integrations are needed in
constructing explicit solutions. However, for \dstwo equation, usual
Darboux transformation without integration is known. This Darboux
transformation is simpler than the binary Darboux transformation for
the hyperbolic NNV equation, and can be easily used to the
stationary \dstwo equation so that explicit solutions of the
hyperbolic NNV equation can be constructed.

Note that the coefficients of $L(\partial)$, $M(\partial)$,
$N(\partial)$ satisfy
\begin{equation}
   \I J,P,Q,R,S\in\Sigma
\end{equation}
where
\begin{equation}
   \Sigma=\{A\hbox{ is a $2\times 2$ matrix}\,|\,KAK^{-1}=\bar A\}
   =\left\{\left(\begin{array}{cc}a&b\\-\bar b&\bar a\end{array}\right)\,
   \Big|\,a,b\in\hc\right\},
\end{equation}
$\nar\D K=\left(\begin{array}{cc}&-1\\1\end{array}\right)$. That is,
$L(\partial)$, $M(\partial)$, $N(\partial)$ satisfy
\begin{equation}
   KL(\partial)K^{-1}=\bar L(\partial),\quad
   KM(\partial)K^{-1}=\bar M(\partial),\quad
   KN(\partial)K^{-1}=\bar N(\partial).\label{eq:reduc_LMN}
\end{equation}
Hence, if
$\nar\D\Phi=\left(\begin{array}{c}\xi\\\eta\end{array}\right)$ is a
solution of (\ref{eq:LP}) with $\lambda=\lambda_0$, then $\nar\D
K\bar\Phi=\left(\begin{array}{c}-\bar\eta\\\bar\xi\end{array}\right)$
is a solution of (\ref{eq:LP}) with $\lambda=\pm\I\bar\lambda_0$.

The Darboux transformation of arbitrary order is constructed as
follows \cite{bib:GHZbook,bib:Zhou}. Suppose
\begin{equation}
   G(\partial)=\partial^n+G_1(x,y,t)\partial^{n-1}+\cdots+G_n(x,y,t)
\end{equation}
is a Darboux operator for (\ref{eq:LP}), i.e., there exist
$L'(\partial)$, $M'(\partial)$, $N'(\partial)$ which have the same
form as $L(\partial)$, $M(\partial)$, $N(\partial)$ with $f$ and $g$
replaced by certain $f'$ and $g'$, such that $\Phi'=G(\partial)\Phi$
satisfies
\begin{equation}
   \lambda\Phi'=L'(\partial)\Phi',\quad
   \Phi'_y=M'(\partial)\Phi',\quad
   \Phi'_t=N'(\partial)\Phi'.
\end{equation}
If so, $G(\partial)$ satisfies
\begin{equation}
   \begin{array}{l}
   L'(\partial)G(\partial)=G(\partial)L(\partial),\\
   M'(\partial)G(\partial)=G(\partial)M(\partial)+G_y(\partial),\\
   N'(\partial)G(\partial)=G(\partial)N(\partial)+G_t(\partial).
   \end{array}\label{eq:DT_hypNNVn}
\end{equation}
Since $L(\partial)$, $M(\partial)$, $N(\partial)$ satisfy the
relations (\ref{eq:reduc_LMN}), and $L'(\partial)$, $M'(\partial)$,
$N'(\partial)$ satisfy the similar relations
\begin{equation}
   KL'(\partial)K^{-1}=\bar{L}'(\partial),\quad
   KM'(\partial)K^{-1}=\bar{M}'(\partial),\quad
   KN'(\partial)K^{-1}=\bar{N}'(\partial),
\end{equation}
we want that $G(\partial)$ satisfies $KG(\partial)K^{-1}=\bar
G(\partial)$. Write
\begin{equation}
   G_j=\left(\begin{array}{cc}a_j &b_j\\-\bar b_j &\bar a_j\end{array}\right).
\end{equation}
Denote $L'(\partial)=2\I J\partial^2+2 P'\partial+Q'$, then the
first equation of (\ref{eq:DT_hypNNVn}) leads to
\begin{equation}
   \begin{array}{l}
   (2\I J\partial^2+2P'\partial+Q')(\partial^n+G_1\partial^{n-1}+\cdots+G_n)\\
   =(\partial^n+G_1\partial^{n-1}+\cdots+G_n)(2\I
   J\partial^2+2P\partial+Q),
   \end{array}
\end{equation}
in which the coefficients of $\partial^{n+1}$ and $\partial^n$ give
\begin{equation}
   \begin{array}{l}
   P'=P-\I[J,G_1],\\
   Q'=Q-2\I[J,G_2]-2[P,G_1]+2\I[J,G_1]G_1+2nP_x-4\I
   JG_{1,x}.
   \end{array}
\end{equation}
Hence, after the action of Darboux transformation,
\begin{equation}
   \begin{array}{l}
   f'=f-2\I b_1,\\
   g'=g-4\I a_{1,x}-2(\bar fb_1-f\bar b_1)-4\I|b_1|^2,\\
   u'=u+8|b_1|^2-4\I(\bar fb_1-f\bar b_1)
   +4(a_{1}+\bar a_{1})_x,\\
   v'=v+8|b_1|^2-4\I(\bar fb_1-f\bar b_1)
   -4\I(a_{1}-\bar a_{1})_x,\\
   w'=w+8|b_1|^2-4\I(\bar fb_1-f\bar b_1)
   +4\I(a_{1}-\bar a_{1})_x.\\
   \end{array}\label{eq:DT_sol_expr}
\end{equation}

Now take $n$ distinct complex numbers $\lambda_1,\cdots,\lambda_n$
with $\lambda_j=\mu_j+\I\nu_j$ ($\mu_j$'s and $\nu_j$'s are real). Let
$\Phi_j=\left(\begin{array}{c}\xi_j\\\eta_j\end{array}\right)$ be a
column solution of (\ref{eq:LP}) with $\lambda=\lambda_j$, then
$\Phi_{n+j}\equiv K\bar
\Phi_j=\left(\begin{array}{c}-\bar\eta_j\\\bar\xi_j\end{array}\right)$
is a solution of (\ref{eq:LP}) with $\lambda=\pm\I\bar\lambda_j$
$(j=1\cdots,n)$. The Darboux transformation is determined by the
system of linear algebraic equations
\begin{equation}
   G(\partial)\Phi_j=0\quad (j=1,\cdots,2n)\label{eq:DT_lin_eq}
\end{equation}
if it has a unique solution \cite{bib:Zhou}.

Denote
\begin{equation}\nar
   \xi=\left(\begin{array}{c}\xi_1\\\vdots\\\xi_n\end{array}\right),\quad
   \eta=\left(\begin{array}{c}\eta_1\\\vdots\\\eta_n\end{array}\right),\quad
   a=\left(\begin{array}{c}a_1\\\vdots\\a_n\end{array}\right),\quad
   b=\left(\begin{array}{c}b_1\\\vdots\\b_n\end{array}\right),
\end{equation}
then (\ref{eq:DT_lin_eq}) becomes
\begin{equation}
   T\left(\begin{array}{c}a\\b\end{array}\right)
   =-\left(\begin{array}{c}\partial^n\xi\\-\partial^n\bar\eta\end{array}\right)
   \label{eq:lineq_for_multi}
\end{equation}
where
\begin{equation}
   T=\left(\begin{array}{cc}A &B\\-\bar B &\bar A\end{array}\right),
\end{equation}
\begin{equation}
   A=\left(\begin{array}{ccc}
   \partial^{n-1}\xi &\cdots &\xi\end{array}\right),\quad
   B=\left(\begin{array}{ccc}\partial^{n-1}\eta &\cdots &\eta\end{array}\right).
\end{equation}
(\ref{eq:DT_lin_eq}) has a unique solution if and only if $\det T\ne
0$.

\section{Expression of soliton solutions}\label{sect:stn}

For zero seed solution $u=v=w=f=g=0$, (\ref{eq:LP}) becomes
\begin{equation}
   \Phi_{xx}=\lambda^2J\Phi,\quad \Phi_y=\I J\Phi_x,\quad \Phi_t=16\I J\Phi_{xxx}
\end{equation}
with $\Phi=(\xi,\eta)^T$. Hence take
\begin{equation}
   \xi_j=\kappa_j^{(1)}(\E{\rho_j^{(1)}+\I\sigma_j^{(1)}}
   +\E{-\rho_j^{(1)}-\I\sigma_j^{(1)}}),\quad
   \eta_j=\kappa_j^{(2)}(\E{\rho_j^{(2)}+\I\sigma_j^{(2)}}
   +\E{-\rho_j^{(2)}-\I\sigma_j^{(2)}})\label{eq:xieta}
\end{equation}
where
\begin{equation}\fl
   \begin{array}{l}
   \rho_j^{(1)}=\re(\lambda_j x+\I \lambda_j y+16\I \lambda_j^3t)+\rho_{j0}^{(1)}
   =\mu_j x-\nu_j y+16(\nu_j^3-3\mu_j^2\nu_j)t+\rho_{j0}^{(1)},\\
   \rho_j^{(2)}=\re(\I \lambda_j x+\lambda_j y-16\lambda_j^3t)+\rho_{j0}^{(2)}
   =-\nu_j x+\mu_j y-16(\mu_j^3-3\mu_j \nu_j^2)t+\rho_{j0}^{(2)},\\
   \sigma_j^{(1)}=\im(\lambda_j x+\I \lambda_j y+16\I \lambda_j^3t)+\sigma_{j0}^{(1)}
   =\nu_j x+\mu_j y+16(\mu_j^3-3\mu_j \nu_j^2)t+\sigma_{j0}^{(1)},\\
   \sigma_j^{(2)}=\im(\I \lambda_j x+\lambda_j y-16\lambda_j^3t)+\sigma_{j0}^{(2)}
   =\mu_j x+\nu_j y+16(\nu_j^3-3\mu_j^2\nu_j)t+\sigma_{j0}^{(2)},
   \end{array}\label{eq:rhosigma}
\end{equation}
$\kappa_j^{(1)}$, $\kappa_j^{(2)}$ are non-zero constants,
$\rho_{j0}^{(1)}$, $\rho_{j0}^{(2)}$, $\sigma_{j0}^{(1)}$,
$\sigma_{j0}^{(2)}$ are real constants. By solving $a_j$'s and
$b_j$'s from (\ref{eq:lineq_for_multi}), the Darboux transformation
(\ref{eq:DT_sol_expr}) gives the $n$-soliton solution
\begin{equation}\fl
   \begin{array}{l}
   f=-2\I b_1,\quad
   g=-4\I a_{1,x}-4\I|b_1|^2,\\
   u=8|b_1|^2+4(a_{1}+\bar a_{1})_x,\quad
   v=8|b_1|^2-4\I(a_{1}-\bar a_{1})_x,\quad
   w=8|b_1|^2+4\I(a_{1}-\bar a_{1})_x.\\
   \end{array}\label{eq:DT_sol_stn}
\end{equation}
Hereafter we omit the primes on $f,g,u,v,w$ for those obtained by
the action of Darboux transformation.

Let $\D K_n=\left(\begin{array}{cc}&-I_n\\I_n\end{array}\right)$.
Denote
\begin{equation}
   \nar\D\zeta=\left(\begin{array}{c}\xi\\-\bar\eta\end{array}\right),\quad
   R_{j\cdots k}=\left(\begin{array}{ccccc}\partial^j\zeta&\partial^{j-1}\zeta,
   \cdots,\partial^{k}\zeta\end{array}\right)
\end{equation}
for $j\ge k$, then
\begin{equation}
   T=\left(\begin{array}{cc}R_{n-1\cdots 0}
   &K_n\bar R_{n-1\cdots 0}
   \end{array}\right).\label{eq:D}
\end{equation}
Let
\begin{equation}\fl
   \begin{array}{l}
   \D\Pi=\left(\begin{array}{ccccccccccccccccc}
   \partial^n\zeta &\partial^{n-1}\zeta &\partial^{n-2}\zeta &R_{n-3\cdots 0}
   &K_n\bar R_{n-1\cdots 0} &0 &0 \\
   \partial^{n+1}\zeta &\partial^{n}\zeta &\partial^{n-1}\zeta
   &0 &0 &R_{n-2\cdots 0} &K_n\bar R_{n-1\cdots 0}\\
   \end{array}\right).
   \end{array}\label{eq:Delta}
\end{equation}

\begin{theorem}
When $\det T\ne 0$, the multi-soliton solution $u$ of the hyperbolic
NNV equation given by (\ref{eq:DT_sol_stn}) can be written as
\begin{equation}
   u=-8\frac{\re\det\Pi}{(\det T)^2}.\label{eq:uexpr}
\end{equation}
\end{theorem}

\begin{demo}
Solved from (\ref{eq:lineq_for_multi}) by Cramer rule,
\begin{equation}\fl
   a_1=-(\det T)^{-1}\left|\begin{array}{ccccccccc}\partial^n\zeta &R_{n-2\cdots 0}
   &K_n\bar R_{n-1\cdots 0}
   \end{array}\right|,
\end{equation}
\begin{equation}\fl
   b_1=-(\det T)^{-1}\left|\begin{array}{ccccccccc}R_{n-1\cdots 0}
   &\partial^{n}\zeta &K_n\bar R_{n-2\cdots 0}
   \end{array}\right|.
\end{equation}
\begin{equation}\fl
   \begin{array}{l}
   \D a_1+\bar a_1\\
   \D=-(\det T)^{-1}\Big(\left|\begin{array}{ccccccccc}
   \partial^n\zeta &R_{n-2\cdots 0} &K_n\bar R_{n-1\cdots 0} \end{array}\right|
   +\overline{\left|\begin{array}{ccccccccc}
   \partial^n\zeta &R_{n-2\cdots 0} &K_n\bar R_{n-1\cdots 0}
   \end{array}\right|}\Big)\\
   \D=-(\det T)^{-1}\Big(\left|\begin{array}{ccccccccc}
   \partial^n\zeta &R_{n-2\cdots 0} &K_n\bar R_{n-1\cdots 0}\end{array}\right|
   +\left|\begin{array}{ccccccccc}R_{n-1\cdots 0}
   &K_n\partial^{n}\bar\zeta &K_n\bar R_{n-2\cdots 0}
   \end{array}\right|\Big)\\
   \D=-(\det T)^{-1} (\det T)_x=-\tr(T^{-1} T_x),
   \end{array}
\end{equation}
\begin{equation}\fl
   a_{1,x}+\bar a_{1,x}=-\tr(T^{-1} T_{xx})+\tr\big((T^{-1}
   T_x)^2\big).
\end{equation}
Denote $\D\nar \widetilde I_k=\left(\begin{array}{c}I_{k\times
k}\\0_{(n-k)\times k}\end{array}\right)$. Let $\nar\D
h=\left(\begin{array}{c}\widetilde a\\\widetilde
b\end{array}\right)$  be the solution of $Th=-\partial^{n+1}\zeta$
where $\widetilde a=(\widetilde a_1,\cdots,\widetilde a_n)^T$,
$\widetilde b=(\widetilde b_1,\cdots,\widetilde b_n)^T$, then
\begin{equation}
   \begin{array}{rl}
   \D a_{1,x}+\bar a_{1,x}=&\D-\tr\left(\begin{array}{cccccc}
   -\widetilde a&-a&\widetilde I_{n-2}&\bar{\widetilde b}&\bar b&0\\
   -\widetilde b&-b&0&-\bar{\widetilde a}&-\bar a&\widetilde I_{n-2}
   \end{array}\right)\\
   &\D+\tr\left(\begin{array}{cccccc}
   -a&\widetilde I_{n-1}&\bar b&0\\
   -b&0&-\bar a&\widetilde I_{n-1}\\
   \end{array}\right)^2\\
   =&a_1^2+\bar a_1^2+\widetilde a_1+\bar{\widetilde a_1}-a_2-\bar a_2-2|b_1|^2.
   \end{array}
\end{equation}
According to (\ref{eq:DT_sol_stn}),
\begin{equation}
   u=8\re(a_1^2+\widetilde a_1-a_2).
\end{equation}
Let $d=-(\det T)^2(a_1^2+\widetilde a_1-a_2)$, then by Cramer rule,
\begin{equation}\fl
   \begin{array}{rl}
   \D d=&\D-\left|\begin{array}{cccccccccc}
   \partial^n\zeta &R_{n-2\cdots 0} &K_n\bar R_{n-1\cdots 0}
   \end{array}\right|^2+\det T\left|\begin{array}{cccccccccc}
   \partial^{n+1}\zeta &R_{n-2\cdots 0} &K_n\bar R_{n-1\cdots 0}
   \end{array}\right|\\
   &\D+\det T\left|\begin{array}{cccccccccc}
   \partial^n\zeta &\partial^{n-1}\zeta &R_{n-3\cdots 0} &K_n\bar R_{n-1\cdots 0}
   \end{array}\right|.
   \end{array}
\end{equation}
Using Laplace expansion of $\det\Pi$, $d=\det\Pi$. Hence
\begin{equation}
   u=-8\frac{\re\,d}{(\det T)^2}=-8\frac{\re\det\Pi}{(\det T)^2}.
\end{equation}
The theorem is proved.
\end{demo}

\begin{remark}
According to Lemma~\ref{lemma:positive} of \ref{appendix}, $\det
T\ge 0$ holds everywhere. However, $\det T>0$ may not hold
everywhere when the parameters $\rho_{j0}^{(k)}$ and
$\sigma_{j0}^{(k)}$ take some special values, as will shown in the
next section for single soliton solution. On the other hand, $\det
T>0$ holds everywhere in generic case, which will be shown here.

The Darboux operator $G(\partial)$ of order $n$ can be constructed
by composing $n$ Darboux operators of order one as follows. For
given $\lambda_1,\cdots,\lambda_n$ and $\Phi_1,\cdots,\Phi_n$ as
above, let $H_j=(\Phi_j,\Phi_{n+j})$ $j(=1,\cdots,n)$. If $\det
H_1\ne 0$, then $\Delta_1(\partial)=\partial-H_{1,x}H_1^{-1}$ is a
Darboux operator of order one. It transforms $(u,v,w,f,g)$ to
$(u^{(1)},v^{(1)},w^{(1)},f^{(1)},g^{(1)})$ and transforms $H_j$ to
$H_j^{(1)}=\Delta_1(\partial)H_j=H_{j,x}-H_{1,x}H_1^{-1} H_j$
$(j=2,3,\cdots,n)$. Again, if $\det H_2^{(1)}\ne 0$, then
$\Delta_2(\partial)=\partial-H_{2,x}^{(1)}(H_2^{(1)})^{-1}$ is a
Darboux operator of order one for the Lax pair with
$(u^{(1)},v^{(1)},w^{(1)},f^{(1)},g^{(1)})$. It trans\-forms
$(u^{(1)},v^{(1)},w^{(1)},f^{(1)},g^{(1)})$ to
$(u^{(2)},v^{(2)},w^{(2)},f^{(2)},g^{(2)})$ and transforms
$H_j^{(1)}$ to
$H_j^{(2)}=\Delta_2(\partial)H_j^{(1)}=H_{j,x}^{(1)}-H_{2,x}^{(1)}(H_2^{(1)})^{-1}
H_j^{(1)}$ $(j=3,4,\cdots,n)$. Continuing this process, we get
$H_j^{(k)}$ $(k=1,\cdots,n-1;\,j=k+1,\cdots,n)$ and
$\Delta_j(\partial)$ $(j=1,\cdots,n)$. According to \cite{bib:Zhou},
\begin{equation}
   \begin{array}{l}
   G(\partial)=\Delta_n(\partial)\Delta_{n-1}(\partial)\cdots\Delta_1(\partial),\\
   \det T=\det(H_n^{(n-1)})\det(H_{n-1}^{(n-2)})\cdots\det(H_2^{(1)})\det(H_1).
   \end{array}
\end{equation}
Hence $\det T\ne 0$ if all $\det H_j^{(j-1)}\ne 0$.

Suppose $\D H_j^{(j-1)}=\left(\begin{array}{cc}\xi_j^{(j-1)}&-\bar\eta_j^{(j-1)}\\
\eta_j^{(j-1)}&\bar\xi_j^{(j-1)}\end{array}\right)$, then $\det
H_j^{(j-1)}=|\xi_j^{(j-1)}|^2+|\eta_j^{(j-1)}|^2=0$ if and only if
$\xi_j^{(j-1)}=0$ and $\eta_j^{(j-1)}=0$ hold simultaneously. For
fixed $j$, this gives a system of four real equations
\begin{equation}
   \begin{array}{l}
   \re\xi_j^{(j-1)}=0,\quad \im\xi_j^{(j-1)}=0,\quad
   \re\eta_j^{(j-1)}=0,\quad \im\eta_j^{(j-1)}=0
   \end{array}
\end{equation}
for three real variables $x,y,t$. It has no solution unless the
parameters $\rho_{j0}^{(k)}$ and $\sigma_{j0}^{(k)}$
$(j=1,\cdots,n;\,k=1,2)$ take special values. This shows that $\det
T>0$ holds everywhere for generic $\rho_{j0}^{(k)}$ and
$\sigma_{j0}^{(k)}$. Therefore, the multi-soliton solution $u$ is global
for generic $\rho_{j0}^{(k)}$ and $\sigma_{j0}^{(k)}$.
\end{remark}

\section{Single soliton solution}\label{sect:1stn}

By taking $n=1$, the single soliton can be obtained as
\begin{equation}
   u=\frac{16A}{B^2}
\end{equation}
where
\begin{equation}\fl
   \begin{array}{rl}
   B=&(\kappa_1^{(1)})^2\cosh(2\rho_1^{(1)})+(\kappa_1^{(2)})^2\cosh
   (2\rho_1^{(2)})+(\kappa_1^{(1)})^2\cos(2\sigma_1^{(1)})
   +(\kappa_1^{(2)})^2\cos(2\sigma_1^{(2)}),\\
   A=&-(\mu_1^2-\nu_1^2)(\kappa_1^{(1)})^4\cosh
   (2\rho_1^{(1)})\cos(2\sigma_1^{(1)})
   -2\mu_1 \nu_1(\kappa_1^{(1)})^4\sinh
   (2\rho_1^{(1)})\sin(2\sigma_1^{(1)})\\
   &-(\mu_1^2+\nu_1^2)(\kappa_1^{(1)})^2(\kappa_1^{(2)})^2\sinh(2\rho_1^{(1)})
   \sin(2\sigma_1^{(2)})\\
   &+(\mu_1^2-\nu_1^2)(\kappa_1^{(2)})^4\cosh(2\rho_1^{(2)})\cos(2\sigma_1^{(2)})
   +2\mu_1 \nu_1(\kappa_1^{(2)})^4\sinh (2\rho_1^{(2)})\sin(2\sigma_1^{(2)})\\
   &+(\mu_1^2+\nu_1^2)(\kappa_1^{(1)})^2(\kappa_1^{(2)})^2\sinh(2\rho_1^{(2)})
   \sin(2\sigma_1^{(1)})\\
   &+(\mu_1^2-\nu_1^2)((\kappa_1^{(2)})^4-(\kappa_1^{(1)})^4).
   \end{array}
\end{equation}

The solution is singular if $B=0$, i.e. $|\xi_1|^2+|\eta_1|^2=0$.
This is equivalent to $\rho_1^{(1)}=\rho_1^{(2)}=0$,
$2\sigma_1^{(1)}=j\pi+\pi/2$, $2\sigma_1^{(2)}=k\pi+\pi/2$ for
certain integers $j$ and $k$. In contrast, the solution is global if
and only if $|\xi_1|^2+|\eta_1|^2\ne0$ everywhere, i.e. the
parameters satisfy
\begin{equation}
   \mu_1(\rho_{10}^{(1)}-\sigma_{10}^{(2)}+k\pi+\pi/2)
   +\nu_1(\rho_{10}^{(2)}+\sigma_{10}^{(1)}-j\pi-\pi/2)\ne 0.\label{eq:1Stn_global_cond}
\end{equation}
We always suppose (\ref{eq:1Stn_global_cond}) is satisfied, which is
equivalent to $\det T\ne 0$.

\begin{figure}
\begin{center}
\includegraphics[240,160]{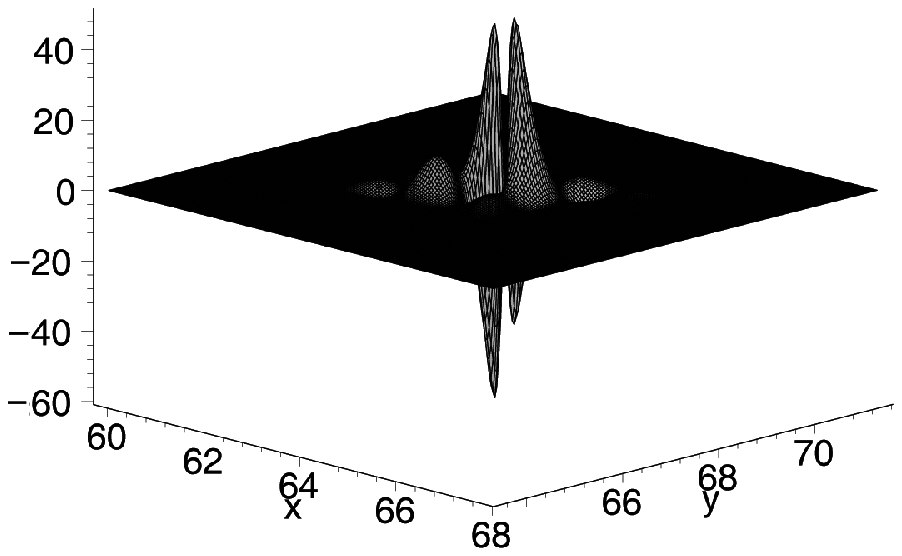}
\caption{Single soliton solution $u$: $\lambda_1=2+0.5\I$,
$\kappa_1^{(1)}=1$, $\kappa_1^{(2)}=1.2$, $\rho_{10}^{(1)}=0$,
$\rho_{10}^{(2)}=1$, $\sigma_{10}^{(1)}=\sigma_{10}^{(2)}=0$,
$t=1$.}\label{fig:NNVsoliton1a}
\end{center}
\end{figure}

\begin{figure}
\begin{center}
\includegraphics[240,160]{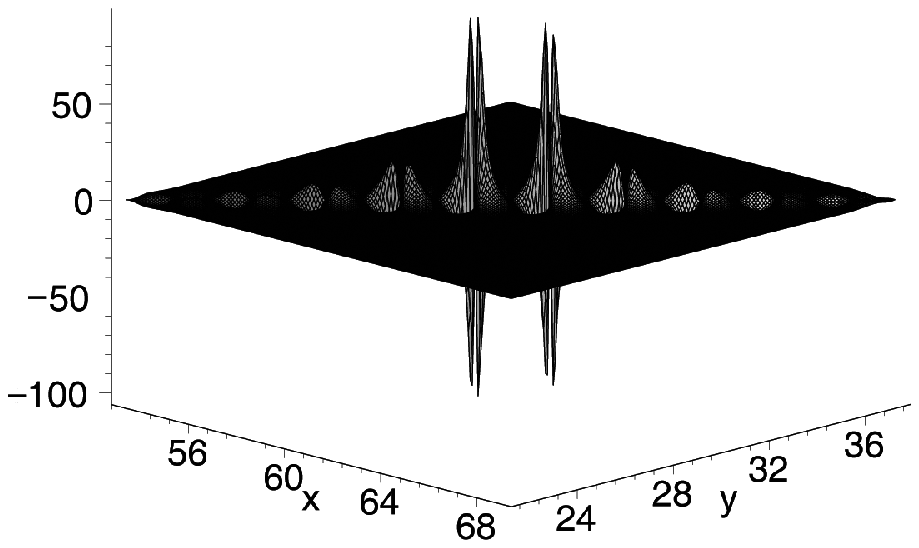}
\caption{Single soliton solution $u$: $\lambda_1=1.1+0.9\I$,
$\kappa_1^{(1)}=1$, $\kappa_1^{(2)}=1.2$, $\rho_{10}^{(1)}=0$,
$\rho_{10}^{(2)}=1$, $\sigma_{10}^{(1)}=\sigma_{10}^{(2)}=0$,
$t=1$.}\label{fig:NNVsoliton1b}
\end{center}
\end{figure}

When $\mu_1^2\ne \nu_1^2$, the solution $u$ approaches zero
exponentially at spatial infinity, and the peaks appear when neither
$\rho_1^{(1)}$ nor $\rho_1^{(2)}$ is large. Hence the center of the
lump of peaks locates near $\rho_1^{(1)}=0$ and $\rho_1^{(2)}=0$,
i.e.,
\begin{equation}
   x=64\mu_1 \nu_1 t-\frac{\mu_1 \rho_{10}^{(1)}+\nu_1 \rho_{10}^{(2)}}{\mu_1^2-\nu_1^2},\quad
   y=16(\mu_1^2+\nu_1^2)t-\frac{\nu_1 \rho_{10}^{(1)}+\mu_1 \rho_{10}^{(2)}}{\mu_1^2-\nu_1^2}.
\end{equation}
The solutions are shown in Figure~\ref{fig:NNVsoliton1a} and
Figure~\ref{fig:NNVsoliton1b} for different parameters. The figure
of the solution contains a lump of peaks rather than a single peak,
and the shape depends on the angle
$\D\arctan\frac{\mu_1^2-\nu_1^2}{2\mu_1 \nu_1}$ between the straight
lines $\rho_1^{(1)}=0$ and $\rho_1^{(2)}=0$. Nevertheless, we still
call it single soliton solution because it is generated from the
zero solution by Darboux transformation, and the peaks in the
solution never separate.

Note that although $u$ is localized, $v$ and $w$ are not.

\begin{figure}
\begin{center}
\includegraphics[240,160]{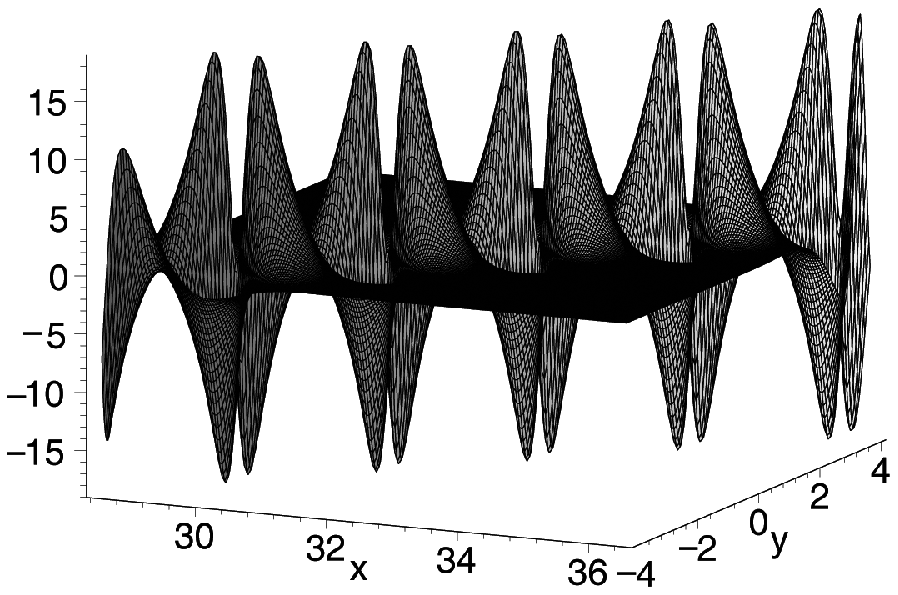}
\caption{Periodic solution $u$: $\lambda_1=1+\I$,
$\kappa_1^{(1)}=1$, $\kappa_1^{(2)}=1.2$, $\rho_{10}^{(1)}=0$,
$\rho_{10}^{(2)}=1$, $\sigma_{10}^{(1)}=\sigma_{10}^{(2)}=0$,
$t=1$.}\label{fig:NNVperiodic1}
\end{center}
\end{figure}

If $\nu_1=\mu_1\ne 0$, the solution is invariant when $(x,y)$ is
changed to $\D\Big(x+\frac{m\pi}{2\mu_1},y+\frac{m\pi}{2\mu_1}\Big)$
for any integer $m$. Hence the solution is periodic. Moreover,
$\rho_1^{(1)}+\rho_1^{(2)}=\rho_{10}^{(1)}+\rho_{10}^{(2)}$. The
peaks appear when neither $\rho_1^{(1)}$ nor $\rho_1^{(2)}$ is
large. Hence the peaks lie near the straight line $\D
x-y-32\mu_1^2t+\frac{\rho_{10}^{(1)}-\rho_{10}^{(2)}}{2\mu_1}=0$.
The solution is shown in Figure~\ref{fig:NNVperiodic1}.

Similarly, the solution is also periodic if $\nu_1=-\mu_1\ne 0$.

\section{Localization of the solutions}\label{sect:nstn}

In this section, we will prove that the multi-soliton solutions
approach zero uniformly and exponentially at spatial infinity. In
order to get global solutions, we always suppose $\det T\ne 0$
everywhere, which is true for generic parameters $\rho_{j0}^{(k)}$ and
$\sigma_{j0}^{(k)}$ $(j=1,\cdots,n;\,k=1,2)$.

Note that the solution of (\ref{eq:lineq_for_multi}) is invariant if
both $\xi_j$ and $\eta_j$ (for fixed $j$) are multiplied by a common
function. Let
\begin{equation}
   \omega_j=\left\{\begin{array}{ll}\xi_j\quad&\hbox{if $|\xi_j|\ge|\eta_j|$}\\
   \eta_j\quad&\hbox{if $|\xi_j|<|\eta_j|$}\end{array}\right.,
\end{equation}
\begin{equation}
   \mathring{T}=\diag(
   \omega_1,\cdots,\omega_n,\bar\omega_1,\cdots,\bar\omega_n).
\end{equation}
Let $\widetilde T=\mathring{T}^{-1}T$, then the norm of each entry
of $\widetilde T$ cannot exceed $1$. Although $\widetilde T$ is not
continuous, $|\det T|$ is continuous.

Let $x=r\cos\theta$, $y=r\sin\theta$. Since $\rho_j^{(1)}$'s and
$\rho_j^{(2)}$'s depend on $x$ and $y$ linearly, we can write, for
$k=1,2$,
\begin{equation}
   \begin{array}{l}
   \rho_j^{(k)}(r\cos\theta,r\sin\theta,t)
   =\varepsilon_j^{(k)}(\theta)\alpha^{(k)}_j(\theta)r+\beta^{(k)}_j,\\
   \end{array}
\end{equation}
where $\varepsilon_j^{(k)}(\theta)=\pm 1$ $(j=1,\cdots,n)$ so that
$\alpha_j^{(k)}(\theta)\ge 0$. Here the variable $t$ is omitted in
$\alpha^{(k)}_j(\theta)$, $\beta^{(k)}_j$ and
$\varepsilon_j^{(k)}(\theta)$.

Clearly $\alpha_j^{(k)}$'s are continuous functions. Note also that
$\varepsilon_j^{(k)}(\theta)$ is not well-defined when
$\alpha_j^{(k)}(\theta)=0$.

\begin{theorem}\label{thm:xy}
Suppose $\lambda_1,\cdots,\lambda_n$ are distinct non-zero complex
numbers such that $\bar\lambda_j\ne\pm\I\lambda_l$ for all
$j,l=1,\cdots,n$. $u$ is the $n$-soliton solution given by
(\ref{eq:uexpr}). Then for fixed $t$, there are positive constants
$r_0$, $\chi$ and $C$ such that
\begin{equation}
   |u(r\cos\theta,r\sin\theta,t)|\le C\E{-\chi r}
\end{equation}
for $r>r_0$ and all $\E{\I\theta}\in S^1$. Hence
$u(r\cos\theta,r\sin\theta,t)\to 0$ uniformly and exponentially as
$r\to+\infty$.
\end{theorem}

\begin{demo}
The proof is divided into four steps.

Step 1: Obtain the asymptotic behavior of $\xi_j$'s and $\eta_j$'s.

Let $\lambda_j=\mu_j+\I\nu_j$ where $\mu_j$'s and $\nu_j$'s are
real, then $\mu_j\ne\pm\nu_j$ for all $j=1,\cdots,n$.

Let $Z^{(\varepsilon)}=\{\E{\I\theta}\in
S^1\,|\,\tan\theta=\varepsilon\}$ for $\varepsilon=\pm 1$,
$Z=Z^{(+1)}\cup Z^{(-1)}$. If $\E{\I\theta}\in Z^{(\varepsilon)}$,
then by (\ref{eq:rhosigma}),
$\D\alpha_j^{(1)}(\theta)=\alpha_j^{(2)}(\theta)=\frac
1{\sqrt2}|\mu_j-\varepsilon\nu_j|>0$,
$\varepsilon_j^{(2)}(\theta)=\varepsilon\varepsilon_j^{(1)}(\theta)$
and
$\varepsilon_j^{(1)}(\theta)(\mu_j-\varepsilon\nu_j)\cos\theta>0$
for all $j=1,\cdots,n$. If $\E{\I\theta}\in S^1\backslash Z$, then
$\alpha_j^{(1)}(\theta)=|\mu_j\cos\theta-\nu_j\sin\theta|$,
$\alpha_j^{(2)}(\theta)=|-\nu_j\cos\theta+\mu_j\sin\theta|$ with
$\alpha_j^{(1)}(\theta)\ne\alpha_j^{(2)}(\theta)$.

For $\delta\in(0,\pi/4)$, define
\begin{equation}
   \begin{array}{l}
   \Omega_\delta^{(\varepsilon)}=\Big\{\E{\I\theta}\,\Big|\,
   \hbox{there exists $\E{\I\theta_0}\in Z^{(\varepsilon)}$
   such that $|\theta-\theta_0|<\delta$}\Big\}\quad
   (\varepsilon=\pm 1),\\
   \Omega_\delta^{(0)}=\Big\{\E{\I\theta}\,\Big|\,
   \hbox{$|\theta-\theta_0|>\delta/2$ for all $\E{\I\theta_0}\in Z$}\Big\}.
   \end{array}
\end{equation}
Then
$\Omega_\delta^{(+1)}\cup\Omega_\delta^{(-1)}\cup\Omega_\delta^{(0)}=S^1$,
and there exists $\delta\in (0,\pi/4)$ and $\omega>0$ such that
\begin{equation}\fl
   \begin{array}{l}
   \alpha_j^{(1)}(\theta)>\omega,\;
   \alpha_j^{(2)}(\theta)>\omega,\quad
   \varepsilon_j^{(2)}(\theta)=\varepsilon\varepsilon_j^{(1)}(\theta),\;
   \varepsilon_j^{(1)}(\theta)(\mu_j-\varepsilon\nu_j)\cos\theta>0\quad
   \hbox{if }\E{\I\theta}\in\Omega_\delta^{(\varepsilon)},\\
   |\alpha_j^{(1)}(\theta)-\alpha_j^{(2)}(\theta)|>\omega\quad\hbox{if
   }\E{\I\theta}\in\Omega_\delta^{(0)}.
   \end{array}
\end{equation}
For $\E{\I\theta_0}\in Z^{(\varepsilon)}$ and
$\E{\I\theta}\in\Omega_\delta^{(\varepsilon)}$ with $|\theta-\theta_0|<\delta$,
$\varepsilon_j^{(1)}(\theta)$ is a constant,
\begin{equation}
   \begin{array}{l}
   \alpha_j^{(1)}(\theta)-\alpha_j^{(2)}(\theta)\\
   =\varepsilon_j^{(1)}(\theta)(\mu_j\cos\theta-\nu_j\sin\theta)
   -\varepsilon_j^{(2)}(\theta)(-\nu_j\cos\theta+\mu_j\sin\theta)\\
   =\varepsilon\varepsilon_j^{(1)}(\theta)(\mu_j+\varepsilon\nu_j)
   \cos\theta(\varepsilon-\tan\theta).
   \end{array}
\end{equation}
Hence, if $\alpha_j^{(1)}(\theta)>\alpha_j^{(2)}(\theta)$ for
$\E{\I\theta}\in\Omega_\delta^{(\varepsilon)}\backslash
Z^{(\varepsilon)}$ with $0<\theta-\theta_0<\delta$, then
$\alpha_j^{(1)}(\theta)<\alpha_j^{(2)}(\theta)$ for
$\E{\I\theta}\in\Omega_\delta^{(\varepsilon)}\backslash
Z^{(\varepsilon)}$ with $-\delta<\theta-\theta_0<0$, and
\textit{vise versa}.

Recall that
\begin{equation}\fl
   \begin{array}{rl}
   \xi_j&=\kappa_j^{(1)}(\E{\varepsilon_j^{(1)}(\theta)\lambda_j x
   +\I\varepsilon_j^{(1)}(\theta)\lambda_j y
   +16\I\varepsilon_j^{(1)}(\theta)\lambda_j^3t}
   +\E{-\varepsilon_j^{(1)}(\theta)\lambda_j x-\I\varepsilon_j^{(1)}(\theta)\lambda_j y
   -16\I\varepsilon_j^{(1)}(\theta)\lambda_j^3t})\\
   &=\kappa_j^{(1)}\E{\alpha_j^{(1)}(\theta)r+\varepsilon_j^{(1)}(\theta)\beta_j^{(1)}
   +\I\varepsilon_j^{(1)}(\theta)\sigma_j^{(1)}(\theta,r)}
   \Big(1+\E{-2\alpha_j^{(1)}(\theta)r-2\varepsilon_j^{(1)}(\theta)\beta_j^{(1)}
   -2\I\varepsilon_j^{(1)}(\theta)\sigma_j^{(1)}(\theta,r)}\Big),\\
   \eta_j&=\kappa_j^{(2)}(\E{\I\varepsilon_j^{(2)}(\theta)\lambda_j x
   +\varepsilon_j^{(2)}(\theta)\lambda_j y
   -16\varepsilon_j^{(2)}(\theta)\lambda_j^3t}
   +\E{-\I\varepsilon_j^{(2)}(\theta)\lambda_j x-\varepsilon_j^{(2)}(\theta)\lambda_j y
   +16\varepsilon_j^{(2)}(\theta)\lambda_j^3t})\\
   &=\kappa_j^{(2)}\E{\alpha_j^{(2)}(\theta)r+\varepsilon_j^{(2)}(\theta)\beta_j^{(2)}
   +\I\varepsilon_j^{(2)}(\theta)\sigma_j^{(2)}(\theta,r)}
   \Big(1+\E{-2\alpha_j^{(2)}(\theta)r-2\varepsilon_j^{(2)}(\theta)\beta_j^{(2)}
   -2\I\varepsilon_j^{(2)}(\theta)\sigma_j^{(2)}(\theta,r)}\Big).\label{eq:recallxieta}
   \end{array}
\end{equation}

Let
\begin{equation}\fl
   Y_j(\theta,r)=(\kappa_j^{(1)})^{-1}\kappa_j^{(2)}
   \E{(\alpha_j^{(2)}(\theta)-\alpha_j^{(1)}(\theta))r
   +\varepsilon_j^{(2)}(\theta)\beta_j^{(2)}-\varepsilon_j^{(1)}(\theta)\beta_j^{(1)}
   +\I\varepsilon_j^{(2)}(\theta)\sigma_j^{(2)}(\theta,r)
   -\I\varepsilon_j^{(1)}(\theta)\sigma_j^{(1)}(\theta,r)}.
   \label{eq:Ydef}
\end{equation}
When $r\to+\infty$, the following limits hold uniformly.

For $\E{\I\theta}\in \Omega_\delta^{(\varepsilon)}$ with
$\alpha_j^{(1)}(\theta)\ge\alpha_j^{(2)}(\theta)$,
\begin{equation}
   \begin{array}{l}
   \D\xi_j^{-1}\partial^k\xi_j\to(\varepsilon_j^{(1)}(\theta)\lambda_j)^k,\quad
   \xi_j^{-1}\partial^k\eta_j-(\I\varepsilon_j^{(2)}(\theta)\lambda_j)^kY_j(\theta,r)\to0.\\
   \end{array}
   \label{eq:asymp_xy_0+}
\end{equation}

For $\E{\I\theta}\in \Omega_\delta^{(\varepsilon)}$ with
$\alpha_j^{(1)}(\theta)\le\alpha_j^{(2)}(\theta)$,
\begin{equation}
   \begin{array}{l}
   \D\eta_j^{-1}\partial^k\xi_j
   -(\varepsilon_j^{(1)}(\theta)\lambda_j)^kY_j(\theta,r)^{-1}\to0,\quad
   \eta_j^{-1}\partial^k\eta_j\to(\I\varepsilon_j^{(2)}(\theta)\lambda_j)^k.
   \end{array}
   \label{eq:asymp_xy_0-}
\end{equation}

For $\E{\I\theta}\in\Omega_\delta^{(0)}$ with
$\alpha_j^{(1)}(\theta)>\alpha_j^{(2)}(\theta)$,
\begin{equation}
   \begin{array}{l}
   \D\xi_j^{-1}\partial^k\xi_j\to(\varepsilon_j^{(1)}(\theta)\lambda_j)^k,\quad
   \D\xi_j^{-1}\partial^k\eta_j\to 0.
   \end{array}\label{eq:asymp_xy_+}
\end{equation}

For $\E{\I\theta}\in\Omega_\delta^{(0)}$ with
$\alpha_j^{(1)}(\theta)<\alpha_j^{(2)}(\theta)$,
\begin{equation}
   \begin{array}{l}
   \D\eta_j^{-1}\partial^k\xi_j\to 0,\quad
   \D\eta_j^{-1}\partial^k\eta_j\to(\I\varepsilon_j^{(2)}(\theta)\lambda_j)^k.
   \end{array}\label{eq:asymp_xy_-}
\end{equation}

Step 2: There exists $r_0>0$ and $c_0>0$ such that $\D\det\widetilde
T>c_0$ when $r>r_0$.

When $\E{\I\theta}\in Z^{(\varepsilon)}$ $(\varepsilon=\pm 1)$,
$\alpha_j^{(1)}(\theta)=\alpha_j^{(2)}(\theta)$ for $j=1,\cdots,n$.
(\ref{eq:recallxieta}) implies
\begin{equation}
   \frac{|\eta_j(\theta,r)|}{|\xi_j(\theta,r)|}\to
   \gamma_j(\theta)\equiv \frac{|\kappa_j^{(2)}|\E{\varepsilon_j^{(2)}(\theta)\beta_j^{(2)}}}
   {|\kappa_j^{(1)}|\E{\varepsilon_j^{(1)}(\theta)\beta_j^{(1)}}}\label{eq:defgammej}
\end{equation}
as $r\to+\infty$. By (\ref{eq:asymp_xy_0+}),
\begin{equation}
   \det\widetilde T(\theta,r)=\Big(\prod_{|\gamma_j(\theta)|>1}|\gamma_j(\theta)|\Big)^{-2}
   \left|\begin{array}{cc}A(\theta,r)&B(\theta,r)
   \\-\bar B(\theta,r)&\bar A(\theta,r)\end{array}\right|+o(1),
\end{equation}
where $A(\theta,r)$ and $B(\theta,r)$ are $n\times n$ matrices,
whose entries are
\begin{equation}\fl
   A_{jk}(\theta,r)=(\varepsilon_j^{(1)}(\theta)\lambda_j)^k,\quad
   B_{jk}(\theta,r)=(\I\varepsilon_j^{(2)}(\theta)\lambda_j)^kY_j(\theta,r)
   =(\I\varepsilon\varepsilon_j^{(1)}(\theta)\lambda_j)^kY_j(\theta,r)
\end{equation}
and $o(1)$ refers to the terms which tend to zero as $r\to+\infty$.
Let
\begin{equation}
   \begin{array}{l}
   \Lambda=(\varepsilon_1^{(1)}(\theta)\lambda_1,\cdots,\varepsilon_n^{(1)}(\theta)\lambda_n),\\
   \Gamma=\diag(Y_1(\theta,r),\cdots,Y_n(\theta,r)).
   \end{array}
\end{equation}
By Lemma~\ref{lemma:bounded_infty_alg} of \ref{appendix}, there
exist $r_1>0$ and $c_1>0$ such that $\D\det\widetilde T>c_1$ for
$\E{\I\theta}\in Z^{(\varepsilon)}$ and $r>r_1$.

When $\E{\I\theta}\in\Omega_\delta^{(\varepsilon)}\backslash Z$
$(\varepsilon=\pm 1)$ with $0<\theta-\theta_0<\delta$ ($\theta_0\in
Z^{(\varepsilon)}$),
$\varepsilon_j^{(2)}(\theta)=\varepsilon\varepsilon_j^{(1)}(\theta)$.
Suppose $\alpha_j^{(1)}(\theta)>\alpha_j^{(2)}(\theta)$ for
$j=1,\cdots,m$ and $\alpha_j^{(1)}(\theta)<\alpha_j^{(2)}(\theta)$
for $j=m+1,\cdots,n$. By (\ref{eq:asymp_xy_0+}) and
(\ref{eq:asymp_xy_0-}),
\begin{equation}\fl
   \begin{array}{l}
   \det\widetilde T(\theta,r)=
   \left|\begin{array}{cc}
   A(\theta)&B(\theta,r)\\C(\theta,r)&D(\theta)\\
   -\bar B(\theta,r)&\bar A(\theta)\\-\bar D(\theta)&\bar C(\theta,r)
   \end{array}\right|+o(1)
   \D=\left|\begin{array}{cc}
   A(\theta)&B(\theta,r)\\\bar D(\theta)&-\bar C(\theta,r)
   \\-\bar B(\theta,r)&\bar A(\theta)\\C(\theta,r)&D(\theta)
   \end{array}\right|+o(1)
   \end{array}\label{eq:detT_xy_vdm}
\end{equation}
where $A(\theta)$ and $B(\theta,r)$ are $m\times n$ matrices,
$C(\theta,r)$ and $D(\theta)$ are $(n-m)\times n$ matrices, whose
entries are given by
\begin{equation}\fl
   \begin{array}{l}
   \D A_{jk}(\theta)=(\varepsilon_j^{(1)}(\theta)\lambda_j)^k,\quad
   B_{jk}(\theta,r)=(\I\varepsilon\varepsilon_j^{(1)}(\theta)\lambda_j)^kY_j(\theta,r)\quad
   (j=1,\cdots,m),\\
   \D\bar D_{jk}(\theta)=(-\I\varepsilon_j^{(2)}(\theta)\bar\lambda_j)^k,\;
   -\bar C_{jk}(\theta,r)
   =(\varepsilon\varepsilon_j^{(2)}(\theta)\bar\lambda_j)^k
   (-\bar Y_j(\theta,r)^{-1})\quad
   (j=m+1,\cdots,n)
   \end{array}
\end{equation}
and $o(1)$ refers to the terms which tend to zero uniformly as
$r\to+\infty$.

Let
\begin{equation}
   \begin{array}{l}
   \Lambda=(\varepsilon_1^{(1)}(\theta)\lambda_1,\cdots,
   \varepsilon_m^{(1)}(\theta)\lambda_m,
   -\I\varepsilon_{m+1}^{(2)}(\theta)\bar\lambda_{m+1},\cdots,
   -\I\varepsilon_n^{(2)}(\theta)\bar\lambda_n),\\
   \Gamma=\diag(Y_1(\theta,r),\cdots,Y_m(\theta,r),
   -\bar Y_{m+1}(\theta,r)^{-1},\cdots,-\bar Y_n(\theta,r)^{-1}).
   \end{array}
\end{equation}
By Lemma~\ref{lemma:bounded_infty_alg} of \ref{appendix}, there
exist $r_2>0$ and $c_2>0$ such that $\D\det\widetilde T>c_2$ for
$\E{\I\theta}\in\Omega_\delta^{(+1)}\cup\Omega_\delta^{(-1)}\backslash
Z$ with $0<\theta-\theta_0<\delta$ and $r>r_2$.

Similarly, when
$\E{\I\theta}\in\Omega_\delta^{(+1)}\cup\Omega_\delta^{(-1)}\backslash
Z$ with $-\delta<\theta-\theta_0<0$ ($\theta_0\in Z$), there exist
$r_3>0$ and $c_3>0$ such that $\D\det\widetilde T>c_3$ for $r>r_3$.

When $\E{\I\theta}\in\Omega_\delta^{(0)}$, suppose
$\alpha_j^{(1)}(\theta)>\alpha_j^{(2)}(\theta)$ for $j=1,\cdots,m$
and $\alpha_j^{(1)}(\theta)<\alpha_j^{(2)}(\theta)$ for
$j=m+1,\cdots,n$, then, by (\ref{eq:asymp_xy_+}) and
(\ref{eq:asymp_xy_-}),
\begin{equation}
   \begin{array}{l}
   \D\lim_{r\to+\infty}\det\widetilde T(\theta,r)=
   \left|\begin{array}{cc}
   A(\theta)&0\\0&D(\theta)\\
   0&\bar A(\theta)\\-\bar D(\theta)&0
   \end{array}\right|
   \D=\left|\begin{array}{cc}
   A(\theta)&0\\\bar D(\theta)&0
   \\0&\bar A(\theta)\\0&D(\theta)
   \end{array}\right|=|\left|\begin{array}{c}
   A(\theta)\\\bar D(\theta)
   \end{array}\right||^2
   \end{array}
\end{equation}
holds uniformly, where $A(\theta)$ is an $m\times n$ matrix,
$D(\theta)$ is an $(n-m)\times n$ matrix, whose entries are given by
\begin{equation}
   \begin{array}{l}
   \D A_{jk}(\theta)=(\varepsilon_j^{(1)}(\theta)\lambda_j)^k,\quad
   (j=1,\cdots,m),\\
   \bar D_{jk}(\theta)=(-\I\varepsilon_j^{(2)}(\theta)\bar\lambda_j)^k\quad
   (j=m+1,\cdots,n).
   \end{array}
\end{equation}
Using the condition $\bar\lambda_j\ne\pm\I\lambda_l$ and the property of
Vandermonde determinant, we know that there exist $r_4>0$ and
$c_4>0$ such that $\D\det\widetilde T>c_4$ for all
$\E{\I\theta}\in\Omega_\delta^{(0)}$ and $r>r_4$.

Let $r_0=\max(r_1,r_2,r_3,r_4)$, $c_0=\min(c_1,c_2,c_3,c_4)$, then
for any $\E{\I\theta}\in S^1$, $\D\det\widetilde T>c_0$ when
$r>r_0$.

Step 3: Denote $\D\nar\widetilde\Pi=\left(\begin{array}{cc}\mathring
T\\&\mathring T\end{array}\right)^{-1}\Pi$, then
$\D\lim_{r\to+\infty}\re\det\widetilde\Pi=0$ for any fixed
$\E{\I\theta}\in S^1$.

When $\E{\I\theta}\in Z^{(\varepsilon)}$, considering
(\ref{eq:asymp_xy_0+}), (\ref{eq:asymp_xy_0-}) and
$\varepsilon_j^{(2)}(\theta)=\varepsilon\varepsilon_j^{(1)}(\theta)$, let
$\Lambda=\diag(\Lambda_1,\Lambda_2)$ with
\begin{equation}
   \begin{array}{l}
   \Lambda_1=\diag(\varepsilon_1^{(1)}(\theta)\lambda_1,\cdots,
   \varepsilon_n^{(1)}(\theta)\lambda_n),\\
   \Lambda_2=\diag(-\I\varepsilon\varepsilon_1^{(1)}(\theta)\bar\lambda_1,\cdots,
   -\I\varepsilon\varepsilon_n^{(1)}(\theta)\bar\lambda_n),
   \end{array}
\end{equation}
\begin{equation}
   \zeta=(\underbrace{1,\cdots,1}_n,
   -\bar\xi_1^{-1}\bar\eta_1,\cdots,-\bar\xi_n^{-1}\bar\eta_n)^T,
\end{equation}
then $\bar\Lambda_2=\I\varepsilon\Lambda_1$, and
$\D\widetilde\Pi-\Big(\prod_{|\gamma_j(\theta)|>1}|\gamma_j(\theta)|\Big)^{-4}\Pi^\Lambda\to
0$ as $r\to+\infty$ where $\gamma_j(\theta)$'s are defined by
(\ref{eq:defgammej}) and $\Pi^\Lambda$ is defined by
(\ref{eq:Delta2}). According to Lemma~\ref{lemma:asymp0} of
\ref{appendix}, $\re\det\Pi^\Lambda\equiv 0$, which leads to
$\D\lim_{r\to+\infty}\re\det\widetilde\Pi=0$.

When $\E{\I\theta}\in S^1\backslash Z$, suppose
$\alpha_j^{(1)}(\theta)>\alpha_j^{(2)}(\theta)$ for $j=1,\cdots,m$
and $\alpha_j^{(1)}(\theta)<\alpha_j^{(2)}(\theta)$ for
$j=m+1,\cdots,n$. By (\ref{eq:asymp_xy_+}) and
(\ref{eq:asymp_xy_-}),
\begin{equation}\fl
   \begin{array}{l}
   \xi_j^{-1}\Big(\partial^k\xi_j-(\varepsilon_j^{(1)}(\theta)\lambda_j)^k\xi_j\Big)\to
   0,\;
   \xi_j^{-1}\Big(\partial^k\eta_j-(\I\varepsilon_j^{(1)}(\theta)\lambda_j)^k\eta_j\Big)\to
   0\quad(j=1,\cdots,m),\\
   \eta_j^{-1}\Big(\partial^k\xi_j-(\varepsilon_j^{(2)}(\theta)\lambda_j)^k\xi_j\Big)\to
   0,\;
   \eta_j^{-1}\Big(\partial^k\eta_j-(\I\varepsilon_j^{(2)}(\theta)\lambda_j)^k\eta_j\Big)\to
   0\quad(j=m+1,\cdots,n)
   \end{array}
   \label{eq:limit_PI_xy}
\end{equation}
as $r\to+\infty$ since $\xi_j^{-1}\eta_j\to 0$ for $j=1,\cdots,m$
and $\eta_j^{-1}\xi_j\to 0$ for $j=m+1,\cdots,n$.

Let $\Lambda=\diag(\Lambda_1,\Lambda_2)$ with
\begin{equation}\fl
   \begin{array}{l}
   \Lambda_1=\diag(\varepsilon_1^{(1)}(\theta)\lambda_1,\cdots,
   \varepsilon_m^{(1)}(\theta)\lambda_m,
   \varepsilon_{m+1}^{(2)}(\theta)\lambda_{m+1},\cdots,
   \varepsilon_n^{(2)}(\theta)\lambda_n),\\
   \Lambda_2=\diag(-\I\varepsilon_1^{(1)}(\theta)\bar\lambda_1,\cdots,
   -\I\varepsilon_m^{(1)}(\theta)\bar\lambda_m,
   -\I\varepsilon_{m+1}^{(2)}(\theta)\bar\lambda_{m+1},\cdots,
   -\I\varepsilon_n^{(2)}(\theta)\bar\lambda_n),
   \end{array}
\end{equation}
\begin{equation}\fl
   \zeta=(\underbrace{1,\cdots,1}_{m},\underbrace{0,\cdots,0}_{n-m},
   \underbrace{0,\cdots,0}_{m},\underbrace{-1,\cdots,-1}_{n-m})^T,
\end{equation}
then $\bar\Lambda_2=\I\Lambda_1$, and $\widetilde\Pi\to\Pi^\Lambda$
as $r\to+\infty$ where $\Pi^\Lambda$ is defined by
(\ref{eq:Delta2}). According to Lemma~\ref{lemma:asymp0} in
\ref{appendix}, we have
$\D\lim_{r\to+\infty}\re\det\widetilde\Pi=\re\det\Pi^\Lambda=0$.

Till now, we have proved that $\det\widetilde T$ has a uniform
positive lower bound for all $\theta$, and
$\D\lim_{r\to+\infty}\frac{\re\det\Pi}{(\det T)^2}=0$ for any fixed
$\theta$.

Step 4: $\D\lim_{r\to+\infty}\frac{\re\det\Pi}{(\det T)^2}=0$
uniformly for all $\E{\I\theta}\in S^1$ as $r\to+\infty$.

Note that $\D \frac{\re\det\Pi}{(\det T)^2}$ is of form
$\D\frac{f(\theta,r)}{g(\theta,r)}$ where
\begin{equation}
   f(\theta,r)=\sum_{j=1}^{m_1}
   \E{\widetilde\alpha_j(\theta)r+\widetilde\gamma_j(\theta)},\quad
   g(\theta,r)=\sum_{j=1}^{m_2}
   \E{\widetilde\beta_j(\theta)r+\widetilde\delta_j(\theta)}
\end{equation}
are real-valued functions of $(\theta,r)$,
$\widetilde\alpha_j(\theta)$, $\widetilde\beta_j(\theta)$,
$\widetilde\gamma_j(\theta)$, $\widetilde\delta_j(\theta)$ are
(complex valued) continuous functions of $\theta$. Let $\D
\widetilde a(\theta)=\max_{1\le j\le
m_1}\re\widetilde\alpha_j(\theta)$, $\D \widetilde
b(\theta)=\max_{1\le j\le m_2}\re\widetilde\beta_j(\theta)$. Since
$\D\lim_{r\to+\infty}\frac{f(\theta,r)}{g(\theta,r)}=0$ for any
fixed $\theta$, the real continuous function $\widetilde
a(\theta)-\widetilde b(\theta)<0$ achieves its maximum $-\chi<0$ on
the compact set $S^1$. We have known that $\det\widetilde T$ has a
uniform positive lower bound as $r\ge r_0$ ($r_0$ is independent of
$\theta$), so has
\begin{equation}
   \sum_{j=1}^{m_2}\E{(\widetilde\beta_j(\theta)-\widetilde b(\theta))r
   +\widetilde\delta_j(\theta)}.
\end{equation}
Hence
\begin{equation}
   \left|\frac{f(\theta,r)}{g(\theta,r)}\right|
   \le\E{-\chi r}\frac{\left|\D\sum_{j=1}^{m_1}
   \E{(\widetilde\alpha_j(\theta)-\widetilde\alpha(\theta))r+\widetilde\gamma_j(\theta)}\right|}
   {\D\sum_{j=1}^{m_2}\E{(\widetilde\beta_j(\theta)-\widetilde b(\theta))r
   +\widetilde\delta_j(\theta)}}\le C\E{-\chi r}
\end{equation}
as $r\ge r_0$ where $C$ is a constant independent of $\theta$. The
theorem is proved.
\end{demo}

\section{Asymptotic behavior of the solutions as
$t\to\infty$}\label{sect:tinfty}

In this section, the asymptotic behavior of the $n$-soliton
solutions as $t\to\infty$ will be discussed. In order to do so, we
consider the problem in a moving frame. Let $x=x_0+\theta_1t$,
$y=y_0+\theta_2t$ where $(\theta_1,\theta_2)$ is the velocity of the
moving frame, and $(x_0,y_0)$ is the coordinate in the moving frame
with this velocity. Then
\begin{equation}
   \begin{array}{l}
   \rho_j^{(1)}(x_0+\theta_1t,y_0+\theta_2t,t)
   =\varepsilon_j^{(1)}\alpha^{(1)}_jt+\beta^{(1)}_j,\\
   \rho_j^{(2)}(x_0+\theta_1t,y_0+\theta_2t,t)
   =\varepsilon_j^{(2)}\alpha^{(2)}_jt+\beta^{(2)}_j
   \end{array}
\end{equation}
where $\varepsilon_j^{(k)}=\pm 1$ $(j=1,\cdots,n;\,k=1,2)$ so that
$\alpha_j^{(k)}\ge 0$. Write $\lambda_j=\mu_j+\I\nu_j$
($j=1,\cdots,n$) where $\mu_j$'s and $\nu_j$'s are real, then
according to (\ref{eq:rhosigma}),
\begin{equation}
   \begin{array}{l}
   \varepsilon_j^{(1)}\alpha^{(1)}_j=\mu_j\theta_1-\nu_j\theta_2+16(\nu_j^3-3\mu_j^2\nu_j),\\
   \varepsilon_j^{(2)}\alpha^{(2)}_j=-\nu_j\theta_1+\mu_j\theta_2-16(\mu_j^3-3\mu_j\nu_j^2),
   \end{array}
\end{equation}
and
\begin{equation}
   \begin{array}{l}
   \varepsilon_j^{(1)}\alpha^{(1)}_j-\varepsilon_j^{(2)}\alpha^{(2)}_j
   =(\mu_j+\nu_j)(\theta_1-\theta_2+16(\mu_j^2-4\mu_j\nu_j+\nu_j^2)),\\
   \varepsilon_j^{(1)}\alpha^{(1)}_j+\varepsilon_j^{(2)}\alpha^{(2)}_j
   =(\mu_j-\nu_j)(\theta_1+\theta_2-16(\mu_j^2+4\mu_j\nu_j+\nu_j^2)).
   \end{array}\label{eq:alphat+-}
\end{equation}

\begin{theorem}\label{thm:t}
Suppose $\lambda_j=\mu_j+\I\nu_j\ne 0$ ($j=1,\cdots,n$) are distinct
complex numbers where $\mu_j$'s and $\nu_j$'s are real numbers, such
that
\begin{equation}
   \begin{array}{l}
   \mu_j\ne\pm\nu_j\quad\hbox{for all $j$},\\
   \mu_j^2+4\mu_j\nu_j+\nu_j^2\ne\mu_l^2+4\mu_l\nu_l+\nu_l^2\quad
   \hbox{for all $j\ne l$},\\
   \mu_j^2-4\mu_j\nu_j+\nu_j^2\ne\mu_l^2-4\mu_l\nu_l+\nu_l^2\quad
   \hbox{for all $j\ne l$}.
   \end{array}\label{eq:lambda_cond}
\end{equation}
$u$ is the $n$-soliton solution given by (\ref{eq:uexpr}). Then for
bounded $(x_0,y_0)$,
$\D\lim_{t\to\infty}u(x_0+\theta_1t,y_0+\theta_2t,t)=0$ except when
\begin{equation}
   \begin{array}{l}
   \theta_1=8(\mu_l^2+4\mu_l\nu_l+\nu_l^2-\mu_j^2+4\mu_j\nu_j-\nu_j^2),\\
   \theta_2=8(\mu_l^2+4\mu_l\nu_l+\nu_l^2+\mu_j^2-4\mu_j\nu_j+\nu_j^2),\\
   (j,l=1,2,\cdots,n).
   \end{array}\label{eq:multi_soliton_velocity}
\end{equation}
Therefore, as $t\to\infty$, $u$ has at most $n\times n$ lumps of peaks
which move in the above velocities $(\theta_1,\theta_2)$
respectively.
\end{theorem}

\begin{demo}

We will always suppose that $(\theta_1,\theta_2)$ does not satisfy
(\ref{eq:multi_soliton_velocity}). Then, by (\ref{eq:alphat+-}),
$\alpha_j^{(1)}\ne 0$ whenever $\alpha_j^{(1)}=\alpha_j^{(2)}$.
Moreover, we only consider the limit $t\to+\infty$. The conclusion
is the same for $t\to-\infty$.

The proof is divided into three steps.

Step 1: Obtain the asymptotic behavior of $\xi_j$'s and $\eta_j$'s.

Suppose $\alpha_j^{(1)}>\alpha_j^{(2)}$ for $j=1,\cdots,m$;
$\alpha_j^{(1)}<\alpha_j^{(2)}$ for $j=m+1,\cdots,p$;
$\alpha_j^{(1)}=\alpha_j^{(2)}\ne 0$ for $j=p+1,\cdots,n$. Then
\begin{equation}\fl
   \begin{array}{l}
   \xi_j^{-1}\Big(\partial^k\xi_j-(\varepsilon_j^{(1)}\lambda_j)^k\xi_j\Big)\to
   0,\quad
   \xi_j^{-1}\Big(\partial^k\eta_j-(\I\varepsilon_j^{(2)}s_j\lambda_j)^k\eta_j\Big)\to
   0\quad(j=1,\cdots,m),\\
   \eta_j^{-1}\Big(\partial^k\xi_j-(\varepsilon_j^{(1)}s_j\lambda_j)^k\xi_j\Big)\to
   0,\quad
   \eta_j^{-1}\Big(\partial^k\eta_j-(\I\varepsilon_j^{(2)}\lambda_j)^k\eta_j\Big)\to
   0\quad(j=m+1,\cdots,p),\\
   \xi_j^{-1}\Big(\partial^k\xi_j-(\varepsilon_j^{(1)}\lambda_j)^k\xi_j\Big)\to
   0,\quad
   \xi_j^{-1}\Big(\partial^k\eta_j-(\I\varepsilon_j^{(2)}\lambda_j)^k\eta_j\Big)\to
   0\quad(j=p+1,\cdots,n)\\
   \end{array}
   \label{eq:limit_PI_t}
\end{equation}
as $r\to+\infty$ where $s_1,\cdots,s_p$ are any constants, since
$\xi_j^{-1}\eta_j\to 0$ for $j=1,\cdots,m$ and $\eta_j^{-1}\xi_j\to
0$ for $j=m+1,\cdots,p$.

Now we prove that $p$ can only take $n-1$ or $n$. If $p\le n-2$,
then
$\varepsilon_j^{(1)}\varepsilon_j^{(2)}=\pm\varepsilon_l^{(1)}\varepsilon_l^{(2)}$
must hold for any $j\ne l$ with $p+1\le j,l\le n$ since both sides
equal $\pm 1$. If
$\varepsilon_j^{(1)}\varepsilon_j^{(2)}=\varepsilon_l^{(1)}\varepsilon_l^{(2)}$,
then
$\varepsilon_j^{(2)}\alpha_j^{(2)}=\varepsilon\varepsilon_j^{(1)}\alpha_j^{(1)}$,
$\varepsilon_l^{(2)}\alpha_l^{(2)}=\varepsilon\varepsilon_l^{(1)}\alpha_l^{(1)}$
hold simultaneously where
$\varepsilon=\varepsilon_j^{(1)}\varepsilon_j^{(2)}$. This
contradicts condition (\ref{eq:lambda_cond}). If
$\varepsilon_j^{(1)}\varepsilon_j^{(2)}=-\varepsilon_l^{(1)}\varepsilon_l^{(2)}$,
then
$\varepsilon_j^{(2)}\alpha_j^{(2)}-\varepsilon\varepsilon_j^{(1)}\alpha_j^{(1)}=0$,
$\varepsilon_l^{(2)}\alpha_l^{(2)}+\varepsilon\varepsilon_l^{(1)}\alpha_l^{(1)}=0$
hold simultaneously where
$\varepsilon=\varepsilon_j^{(1)}\varepsilon_j^{(2)}$. This
contradicts the assumption that $(\theta_1,\theta_2)$ does not
satisfy (\ref{eq:multi_soliton_velocity}). Hence only $p=n$ or
$p=n-1$ is possible.

Step 2: There exists $t_0>0$ and $c>0$ such that $\D\det\widetilde
T>c$ for $t\ge t_0$.

When $p=n-1$,
\begin{equation}
   \frac{|\xi_n(\theta,r)|}{\max(|\xi_n(\theta,r)|,|\eta_n(\theta,r)|)}\to\gamma_0\equiv
   \frac{|\kappa_n^{(1)}|\E{\varepsilon_n^{(1)}\beta_n^{(1)}}}
   {\max(|\kappa_n^{(1)}|\E{\varepsilon_n^{(1)}\beta_n^{(1)}},
   |\kappa_n^{(2)}|\E{\varepsilon_n^{(2)}\beta_n^{(2)}})}.
   \label{eq:rho0}
\end{equation}
Denote
\begin{equation}
   V(\lambda_j,\cdots,\lambda_l)=\left(\begin{array}{ccc}
   \lambda_j^{n-1}&\cdots&1\\\vdots&&\vdots\\\lambda_l^{n-1}&\cdots&1
   \end{array}\right)_{(l-j+1)\times n}
\end{equation}
for $j\le l$, then, for $p=n-1$, (\ref{eq:limit_PI_t}) leads to
\begin{equation}\fl
   \begin{array}{l}
   \D\det\widetilde T=\gamma_0^2
   \left|\begin{array}{cc}
   V(\varepsilon_1^{(1)}\lambda_1,\cdots,\varepsilon_m^{(1)}\lambda_m) &0\\
   0 &\hskip-12pt
   V(\I\varepsilon_{m+1}^{(2)}\lambda_{m+1},\cdots,\I\varepsilon_{n-1}^{(2)}\lambda_{n-1})\\
   V(\varepsilon_n^{(1)}\lambda_n)
   &\hskip-12pt\xi_n^{-1}\eta_n
   V(\I\varepsilon_n^{(2)}\lambda_n)\\
   0 &V(\varepsilon_1^{(1)}\bar\lambda_1,\cdots,\varepsilon_m^{(1)}\bar\lambda_m)\\
   -V(-\I\varepsilon_{m+1}^{(2)}\bar\lambda_{m+1},\cdots,
   -\I\varepsilon_{n-1}^{(2)}\bar\lambda_{n-1})
   &0\\
   -\bar\xi_n^{-1}\bar\eta_nV(-\I\varepsilon_n^{(2)}\bar\lambda_n)
   &V(\varepsilon_n^{(1)}\bar\lambda_n)\\
   \end{array}\right|+o(1)\\
   \D=\gamma_0^2
   \left|\begin{array}{cc}
   V(\varepsilon_1^{(1)}\lambda_1,\cdots,\varepsilon_m^{(1)}\lambda_m) &0\\
   V(-\I\varepsilon_{m+1}^{(2)}\bar\lambda_{m+1},\cdots,
   -\I\varepsilon_{n-1}^{(2)}\bar\lambda_{n-1}) &0\\
   V(\varepsilon_n^{(1)}\lambda_n)
   &\xi_n^{-1}\eta_n
   V(\I\varepsilon_n^{(2)}\lambda_n)\\
   0 &V(\varepsilon_1^{(1)}\bar\lambda_1,\cdots,\varepsilon_m^{(1)}\bar\lambda_m)\\
   0 &V(\I\varepsilon_{m+1}^{(2)}\lambda_{m+1},\cdots,\I\varepsilon_{n-1}^{(2)}\lambda_{n-1})\\
   -\bar\xi_n^{-1}\bar\eta_nV(-\I\varepsilon_n^{(2)}\bar\lambda_n)
   &V(\varepsilon_n^{(1)}\bar\lambda_n)\\
   \end{array}\right|+o(1)
   \end{array}
\end{equation}
as $t\to+\infty$. Let
\begin{equation}
   \begin{array}{l}
   \Lambda=\diag(\varepsilon_1^{(1)}\lambda_1,\cdots,\varepsilon_m^{(1)}\lambda_m,
   -\I\varepsilon_{m+1}^{(2)}\bar\lambda_{m+1},\cdots,
   -\I\varepsilon_{n-1}^{(2)}\bar\lambda_{n-1},
   \varepsilon_n^{(1)}\lambda_n),\\
   \Gamma=\diag(0,\cdots,0,\xi_n^{-1}\eta_n),\quad
   \varepsilon=\varepsilon_n^{(1)}\varepsilon_n^{(2)},
   \end{array}
\end{equation}
then we get $\D\liminf_{t\to+\infty}\det\widetilde T>0$ by
Lemma~\ref{lemma:bounded_infty_alg} of \ref{appendix}.

Similarly, when $p=n$,
\begin{equation}\fl
   \begin{array}{l}
   \D\lim_{t\to+\infty}\det\widetilde T=
   \left|\begin{array}{cc}
   V(\varepsilon_1^{(1)}\lambda_1,\cdots,\varepsilon_m^{(1)}\lambda_m) &0\\
   V(-\I\varepsilon_{m+1}^{(2)}\bar\lambda_{m+1},\cdots,-\I\varepsilon_n^{(2)}\bar\lambda_n) &0\\
   0 &V(\varepsilon_1^{(1)}\bar\lambda_1,\cdots,\varepsilon_m^{(1)}\bar\lambda_m)\\
   0 &V(\I\varepsilon_{m+1}^{(2)}\lambda_{m+1},\cdots,\I\varepsilon_n^{(2)}\lambda_n)\\
   \end{array}\right|.\\
   \D=\Big|\det
   V(\varepsilon_1^{(1)}\lambda_1,\cdots,\varepsilon_m^{(1)}\lambda_m,
   -\I\varepsilon_{m+1}^{(2)}\bar\lambda_{m+1},\cdots,-\I\varepsilon_n^{(2)}\bar\lambda_n)
   \Big|^2.   \end{array}
\end{equation}
Using the condition $\bar\lambda_j\ne\pm\I\lambda_l$, we get
$\D\liminf_{t\to+\infty}\det\widetilde T>0$ since the Vandermonde
determinant is non-zero.

Step 3: Denote $\D\nar\widetilde\Pi=\left(\begin{array}{cc}\mathring
T\\&\mathring T\end{array}\right)^{-1}\Pi$. If the velocity
$(\theta_1,\theta_2)$ does not satisfy
(\ref{eq:multi_soliton_velocity}), then
$\D\lim_{t\to+\infty}\re\det\widetilde\Pi=0$.

From (\ref{eq:limit_PI_t}), let $\Lambda=\diag(\Lambda_1,\Lambda_2)$
with
\begin{equation}\fl
   \begin{array}{l}
   \Lambda_1=\diag(\varepsilon_1^{(1)}\lambda_1,\cdots,\varepsilon_m^{(1)}\lambda_m,
   \varepsilon_{m+1}^{(1)}s_{m+1}\lambda_{m+1},\cdots,\varepsilon_p^{(1)}s_p\lambda_p,
   \varepsilon_{p+1}^{(1)}\lambda_{p+1},\cdots,\varepsilon_n^{(1)}\lambda_n),\\
   \Lambda_2=\diag(-\I\varepsilon_1^{(2)}\bar s_1\bar\lambda_1,\cdots,-\I\varepsilon_m^{(2)}
   \bar s_m\bar\lambda_m,
   -\I\varepsilon_{m+1}^{(2)}\bar\lambda_{m+1},\cdots,-\I\varepsilon_p^{(2)}\bar\lambda_p,\\
   \qquad-\I\varepsilon_{p+1}^{(2)}\bar\lambda_{p+1},\cdots,-\I\varepsilon_n^{(2)}\bar\lambda_n),
   \end{array}
\end{equation}
\begin{equation}\fl
   \begin{array}{l}
   \zeta=(\underbrace{1,\cdots,1}_{m},\underbrace{0,\cdots,0}_{p-m},
   \underbrace{1,\cdots,1}_{n-p},\underbrace{0,\cdots,0}_{m},
   \underbrace{-1,\cdots,-1}_{p-m},
   -\bar\xi_{p+1}^{-1}\bar\eta_{p+1},\cdots,-\bar\xi_n^{-1}\bar\eta_n)^T,
   \end{array}
\end{equation}
then $\widetilde\Pi-\gamma_0^4\Pi^\Lambda\to 0$ for $p=n-1$ and
$\widetilde\Pi-\Pi^\Lambda\to 0$ for $p=n$ as $t\to+\infty$ where
$\gamma_0$ is defined by (\ref{eq:rho0}) and $\Pi^\Lambda$ is
defined by (\ref{eq:Delta2}). $\bar\Lambda_2
=\I\varepsilon\Lambda_1$ holds for $\varepsilon=\pm 1$ if and only
if $s_j=\varepsilon\varepsilon_j^{(1)}\varepsilon_j^{(2)}$ for
$j=1,\cdots,p$, and
$\varepsilon_j^{(2)}=\varepsilon\varepsilon_j^{(1)}$ for
$j=p+1,\cdots,n$. Since $s_j$ ($j=1,\cdots,p$) can be arbitrary,
$\varepsilon$ can be taken as $\pm 1$ arbitrarily, and $p=n$ or
$n-1$, we have $\bar\Lambda_2 =\I\varepsilon\Lambda_1$ by taking
$\varepsilon=\varepsilon_n^{(1)}\varepsilon_n^{(2)}$ for $p=n-1$,
and either $\varepsilon=1$ or $\varepsilon=-1$ for $p=n$.

According to Lemma~\ref{lemma:asymp0} of \ref{appendix},
$\re\det\Pi^\Lambda\equiv 0$. Hence
$\D\lim_{r\to+\infty}\re\det\widetilde\Pi=0$ if
$(\theta_1,\theta_2)$ does not satisfy
(\ref{eq:multi_soliton_velocity}). The theorem is proved.
\end{demo}

\begin{remark}
If $(\theta_1,\theta_2)$ satisfies
(\ref{eq:multi_soliton_velocity}), then
$\alpha_j^{(1)}=\alpha_j^{(2)}$, $\alpha_l^{(1)}=\alpha_l^{(2)}$,
$\varepsilon_j^{(2)}=\varepsilon_j^{(1)}$,
$\varepsilon_l^{(2)}=-\varepsilon_l^{(1)}$ for $j\ne l$ with $p+1\le
j,l\le n$. Hence there is no common $\varepsilon=\pm 1$ such that
$\varepsilon_i^{(2)}=\varepsilon\varepsilon_i^{(1)}$ holds for all
$i=p+1,\cdots,n$, which contradicts the condition $\bar\Lambda_2
=\I\varepsilon\Lambda_1$ in Lemma~\ref{lemma:asymp0} of
\ref{appendix}. In fact, the solution does not tend to zero in this
case, which can be seen in the following example.
\end{remark}

\begin{figure}
\begin{center}
\includegraphics[240,180]{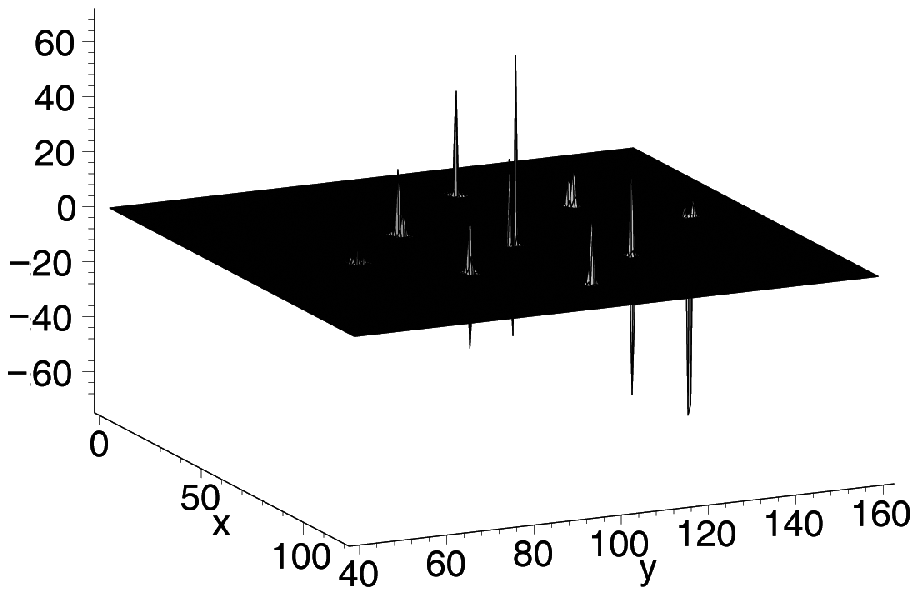}
\caption{$3\times 3$ soliton solution $u$: $\lambda_1=2+0.5\I$,
$\lambda_2=2.5+0.4\I$, $\lambda_3=3+0.3\I$, $\kappa_j^{(1)}=1$,
$\kappa_j^{(2)}=1.2$, $\rho_{j0}^{(1)}=0$, $\rho_{j0}^{(2)}=0$,
$\sigma^{(1)}_{j0}=0$, $\sigma^{(2)}_{j0}=0$ $(j=1,2,3)$,
$t=1$.}\label{fig:NNVSoliton3a}
\end{center}
\end{figure}

\begin{figure}
\begin{center}
\includegraphics[240,160]{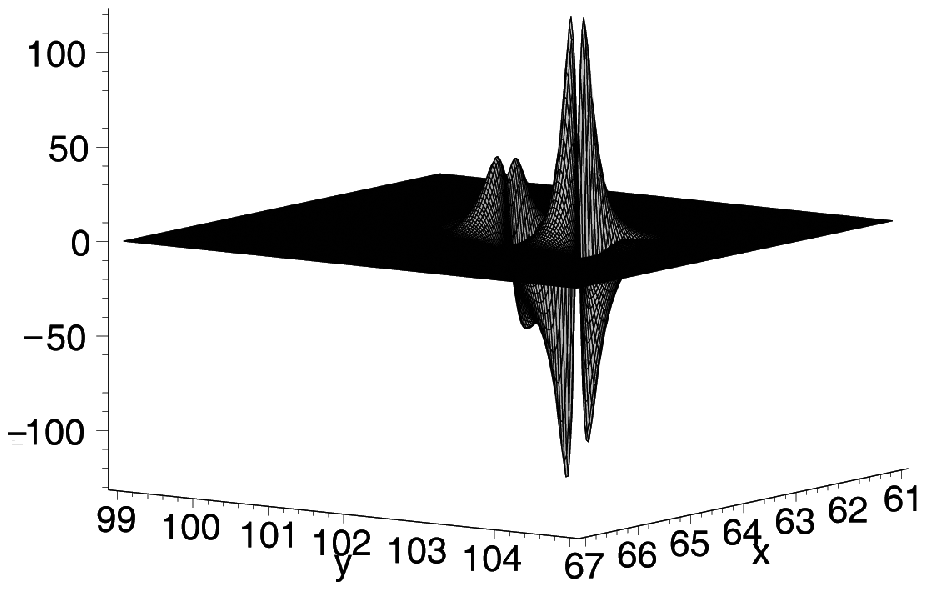}
\caption{Local behavior of one lump of peaks in the $3\times 3$
soliton solution $u$.}\label{fig:NNVSoliton3b}
\end{center}
\end{figure}

As an example, a $3\times 3$ soliton is shown in
Figure~\ref{fig:NNVSoliton3a}, in which there are $9$ lumps of
peaks. The local behavior of each lump of peaks is still complicated
and one of which is shown in Figure~\ref{fig:NNVSoliton3b}.


\appendix

\section{Some linear algebraic lemmas}\label{appendix}

\begin{lemma}\label{lemma:conj}
Suppose $X$ and $Y$ are $2n\times r$ and $2n\times (2n-r)$ matrices
respectively, then
\begin{equation}
   \overline{\left|\begin{array}{cc}X&K_n\bar Y\end{array}\right|}=
   \left|\begin{array}{cc}Y&K_n\bar X\end{array}\right|
\end{equation}
where $\nar\D
K_n=\left(\begin{array}{cc}0&-I_n\\I_n&0\end{array}\right)$.
\end{lemma}

\begin{demo}
\begin{equation}
   K_n\overline{\left(\begin{array}{cc}X&K_n\bar Y\end{array}\right)}K_n^{-1}
   =\left(\begin{array}{cc}K_n\bar X &-Y\end{array}\right)
   \left(\begin{array}{cc}0&I_n\\-I_n&0\end{array}\right)
   =\left(\begin{array}{cc}Y &K_n\bar X\end{array}\right).
\end{equation}
The lemma is obtained by taking the determinants on both sides.
\end{demo}

\begin{lemma}\label{lemma:positive} Suppose $\D
T=\left(\begin{array}{cc}A&B\\-\bar B&\bar A\end{array}\right)$
where $A$ and $B$ are $n\times n$ matrices, then $\det T\ge 0$.
\end{lemma}

\begin{demo} By Lemma~\ref{lemma:conj}, $\det T$ is real.
First suppose both $A$ and $B$ are invertible, then
\begin{equation}
   \det T=\det A\det(\bar A+\bar BA^{-1} B)=|\det
   A|^2\det(I+\overline{A^{-1} B}A^{-1} B).
\end{equation}
Let $N=A^{-1} B$. Suppose $\lambda$ is an eigenvalue of $\bar NN$,
$v\in\hr^{2n}$ is a vector in the corresponding root space
$\RRRR_\lambda$, i.e. $(\bar NN-\lambda I)^mv=0$ for certain positive
integer $m$. Then
\begin{equation}
   (\bar NN-\bar\lambda I)^m(\bar N\bar v)
   =\bar N\overline{(\bar NN-\lambda I)^mv}=0.
\end{equation}
Hence $\bar N\bar v\in\RRRR_{\bar\lambda}$. If $\lambda$ is a
non-real eigenvalue of $\bar NN$ of multiplicity $k$, the
multiplicity of $\bar\lambda$ is also $k$.

Now suppose $\lambda<0$ is an eigenvalue of $\bar NN$,
$\RRRR_\lambda=V_1\oplus\cdots\oplus V_m$ where $V_1,\cdots,V_m$ are
irreducible invariant subspaces. Suppose $V_1=\Span\{\zeta,(\bar
NN-\lambda I)\zeta,\cdots,(\bar NN-\lambda I)^{m-1}\zeta\}$ with
$(\bar NN-\lambda I)^m\zeta=0$. Then $\zeta\not\in\Image(\bar
NN-\lambda I)$, and $(\bar NN-\lambda I)^m\bar N\bar\zeta=\bar
N\overline{(\bar NN-\lambda I)^m\zeta}=0$, $(\bar NN-\lambda
I)^{m-1}\bar N\bar\zeta=\bar N\overline{(\bar NN-\lambda
I)^{m-1}\zeta}\ne 0$ since $\det N\ne 0$. We will prove that
\begin{equation}\fl
   \zeta,(\bar NN-\lambda I)\zeta,\cdots,
   (\bar NN-\lambda I)^{m-1}\zeta,
   \bar N\bar\zeta,(\bar NN-\lambda I)\bar N\bar\zeta,\cdots,
   (\bar NN-\lambda I)^{m-1}\bar N\bar\zeta
   \label{eq:indepvec}
\end{equation}
are linearly independent. Suppose
\begin{equation}
   \sum_{j=1}^m\alpha_j(\bar NN-\lambda I)^{j-1}\zeta+\sum_{j=1}^m\beta_j(\bar NN-\lambda
   I)^{j-1}\bar N\bar\zeta=0\label{eq:lindepzeta}
\end{equation}
where $\alpha_1,\cdots,\alpha_m,\beta_1,\cdots,\beta_m$ are complex
numbers. Acting $(\bar NN-\lambda I)^{m-1}$ on both sides of (\ref{eq:lindepzeta}),
we get
\begin{equation}
   (\bar NN-\lambda I)^{m-1}(\alpha_1\zeta+\beta_1\bar N\bar\zeta)=0.
\end{equation}
Then
\begin{equation}
   \begin{array}{l}
   (|\alpha_1|^2-\lambda|\beta_1|^2)(\bar NN-\lambda I)^{m-1}\zeta\\
   =-\bar\alpha_1\beta_1(\bar NN-\lambda I)^{m-1}\bar N\bar\zeta
   -\lambda|\beta_1|^2(\bar NN-\lambda I)^{m-1}\zeta\\
   =-\bar\alpha_1\beta_1\bar N\overline{(\bar NN-\lambda
   I)^{m-1}\zeta}
   -\lambda|\beta_1|^2(\bar NN-\lambda I)^{m-1}\zeta\\
   =\beta_1\bar N\overline{\beta_1(\bar NN-\lambda
   I)^{m-1}\bar N\bar\zeta}
   -\lambda|\beta_1|^2(\bar NN-\lambda I)^{m-1}\zeta\\
   =|\beta_1|^2\bar NN(\bar NN-\lambda I)^{m-1}\zeta
   -\lambda|\beta_1|^2(\bar NN-\lambda I)^{m-1}\zeta\\
   =|\beta_1|^2(\bar NN-\lambda I)^m\zeta=0.
   \end{array}
\end{equation}
Since $\lambda<0$ and $(\bar NN-\lambda I)^{m-1}\zeta\ne 0$, we have
$\alpha_1=\beta_1=0$. Continuing this process by acting $(\bar
NN-\lambda I)^{m-2}$, $\cdots$, $(\bar NN-\lambda I)^0$ on both
sides of (\ref{eq:lindepzeta}) respectively, we get
$\alpha_1=\cdots=\alpha_n=\beta_1=\cdots=\beta_n=0$. This proves the
linear independence of the vectors in (\ref{eq:indepvec}). Let
$\widetilde V_1=\Span\{\bar N\bar\zeta,(\bar NN-\lambda I)\bar
N\bar\zeta,\cdots,(\bar NN-\lambda I)^{m-1}\bar N\bar\zeta\}$. If
$\bar N\bar\zeta=(\bar NN-\lambda I)\zeta'\in\Image(\bar NN-\lambda
I)$, then $\zeta=(\bar NN-\lambda I)N^{-1}\bar\zeta'\in\Image(\bar
NN-\lambda I)$, which contradicts the choice of $\zeta$. Hence $\bar
N\bar\zeta\not\in\Image(\bar NN-\lambda I)$. Moreover, $\widetilde
V_1$ is invariant and irreducible under the action of $\bar
NN-\lambda I$. Hence it must be one of $V_j$ with $2\le j\le m$,
which means that $m$ is even $(m=2k)$ and
$\RRRR_\lambda=(W_1\oplus\widetilde
W_1)\oplus\cdots\oplus(W_k\oplus\widetilde W_k)$ where
$(W_1,\cdots,W_k,\widetilde W_1,\cdots,\widetilde W_k)$ is a
permutation of $(V_1,\cdots,V_{2k})$. Therefore, the multiplicity of
each negative eigenvalue of $\bar NN$ must be even.

Thus, if the eigenvalues of $\bar NN$ are
$\lambda_1,\cdots,\lambda_{2n}$ (multiple eigenvalues are listed
repeatedly), then
\begin{equation}
   \begin{array}{l}
   \det T=|\det A|^2\det(I+\bar NN)\\
   \D=|\det A|^2\Big(\prod_{\smallre\lambda_j\ne 0}|1+\lambda_j|^2
   \prod_{\lambda_j<0}(1+\lambda_j)^2\Big)^{1/2}
   \prod_{\lambda_j\ge 0}(1+\lambda_j)\ge 0.
   \end{array}
\end{equation}
If $A$ or $B$ is not invertible, it is a limit of invertible
matrices, and the conclusion is also true. The lemma is proved.
\end{demo}

\begin{lemma}\label{lemma:bounded_infty_alg}
Suppose $\Lambda=\diag(\lambda_1,\cdots,\lambda_n)$ where
$\lambda_1,\cdots,\lambda_n$ are distinct complex numbers such that
$\bar{\lambda}_j\ne\pm\I\lambda_l$ for all $j,l=1,\cdots,n$,
$\Gamma=\diag(\gamma_1,\gamma_2,\cdots,\gamma_n)$. Denote
\begin{equation}
   V=\left(\begin{array}{cccc}
   \lambda_1^{n-1}&\lambda_1^{n-2}&\cdots&1\\\vdots&\vdots&&\vdots\\
   \lambda_n^{n-1}&\lambda_n^{n-2}&\cdots&1
   \end{array}\right),\quad
   E=\diag\Big((\I\varepsilon)^{n-1},(\I\varepsilon)^{n-2},\cdots,1\Big)
\end{equation}
where $\varepsilon=\pm 1$. Let
\begin{equation}
   T=\left(\begin{array}{cc} V &\Gamma VE\\
   -\bar \Gamma\bar V\bar E &\bar V
   \end{array}\right),
\end{equation}
then there is a positive number $C$ depending on $\Lambda$ only,
such that $\det T>C$.
\end{lemma}

\begin{demo}
Denote
\begin{equation}
   S_k^{(\hat m)}=\sum_{j_1<\cdots<j_k\atop j_1,\cdots,j_k\ne m}
   \lambda_{j_1}\cdots\lambda_{j_k},
\end{equation}
then
\begin{equation}
   \sum_{k=1}^n(-1)^{k-1}S_{k-1}^{(\hat m)}x^{n-k}
   =\prod_{s=1\atop s\ne m}^n(x-\lambda_s).
\end{equation}
Since the entries of $V$ are $V_{jk}=\lambda_j^{n-k}$, we have
\begin{equation}
   (V^{-1})_{jk}=\frac{(-1)^{j-1}S_{j-1}^{(\hat k)}}
   {\D\prod_{s=1\atop s\ne k}^n(\lambda_k-\lambda_s)}.
\end{equation}
Hence
\begin{equation}
   \begin{array}{l}
   \D\det T
   =\det(V)\det(\bar V+\bar \Gamma\bar V\bar EV^{-1}\Gamma VE)\\
   =|\det(V)|^2\det(I+\bar \Gamma\bar V\bar EV^{-1}\Gamma VE\bar V^{-1}).
   \end{array}
\end{equation}

For $j=1,\cdots,n$, let $\lambda_j=\E{\I\pi/4}\delta\, r_j$ where
$\delta=1$ if $\varepsilon=1$ and $\delta=\I$ if $\varepsilon=-1$,
then $r_j$'s are distinct and $\bar r_j\pm r_l\ne 0$ for all
$j,l=1,\cdots,n$.
\begin{equation}
   \begin{array}{l}
   \D (\bar V\bar EV^{-1})_{jk}
   =\sum_{l=1}^n(-\I\varepsilon\bar{\lambda}_j)^{n-l}
   \frac{(-1)^{l-1}S_{l-1}^{(\hat k)}}{\D\prod_{s=1\atop s\ne k}^n
   (\lambda_k-\lambda_s)}
   =\prod_{s=1\atop s\ne k}^n\frac{-\I\varepsilon\bar{\lambda}_j-\lambda_s}
   {\lambda_k-\lambda_s}\\
   \D=(-1)^{n-1}\prod_{s=1\atop s\ne k}^n\frac{\bar r_j+r_s}{r_k-r_s}
   =(-1)^{n-1}(AMB^{-1})_{jk}
   \end{array}
\end{equation}
where
\begin{equation}
   A=(a_j\delta_{jk}),\quad B=(b_j\delta_{jk}),\quad M=(M_{jk}),
\end{equation}
\begin{equation}
   \begin{array}{l}
   \D a_j=\prod_{s=1}^n(\bar r_j+r_s),\quad b_k=\prod_{s=1\atop s\ne
   k}^n(r_k-r_s),\quad
   \D M_{jk}=(\bar r_j+r_k)^{-1}.
   \end{array}
\end{equation}

Since $\overline{\bar V\bar EV^{-1}}\bar V\bar EV^{-1}=I$, we have
$\overline{AMB^{-1}}=(AMB^{-1})^{-1}$. Hence
\begin{equation}
   \begin{array}{l}
   \D\det T
   =|\det(V)|^2\det(I+\bar \Gamma AMB^{-1}\Gamma BM^{-1} A^{-1})\\
   =|\det(V)|^2(\det M)^{-1}\det(M+\bar \Gamma M\Gamma)
   \end{array}
\end{equation}
since $\Gamma$, $A$, $B$ are diagonal matrices.

Suppose $\re(r_1),\cdots,\re(r_m)<0$,
$\re(r_{m+1}),\cdots,\re(r_n)>0$ since $\bar r_j+r_j\ne 0$ for all
$j$. Write
\begin{equation}
   \begin{array}{l}
   \D M=\Bigg(\frac{1}{\bar
   r_j+r_k}\Bigg)_{n\times n}
   =\left(\begin{array}{cc}M_{11} &M_{12}\\M_{12}^*
   &M_{22}\end{array}\right),\\
   \D M+\bar\Gamma M\Gamma=\Bigg(\frac{1+\bar\gamma_j\gamma_k}{\bar
   r_j+r_k}\Bigg)_{n\times n}
   =\left(\begin{array}{cc}N_{11} &N_{12}\\N_{12}^* &N_{22}\end{array}\right)
   \end{array}
\end{equation}
where $M_{11}$ and $N_{11}$, $M_{12}$ and $N_{12}$, $M_{22}$ and
$N_{22}$ are $m\times m$, $m\times(n-m)$, $(n-m)\times(n-m)$
matrices respectively. Then $M_{11}$ and $N_{11}$ are negative
definite Hermitian matrices, so are $M_{11}^{-1}$ and $N_{11}^{-1}$;
$M_{22}$ and $N_{22}$ are positive definite Hermitian matrices.

Let
\begin{equation}
   \Gamma_1=\diag(\gamma_1,\cdots,\gamma_m),\quad \Gamma_2=\diag(\gamma_{m+1},\cdots,\gamma_n),
\end{equation}
then
\begin{equation}
   \begin{array}{l}
   \det(-N_{11})=\det(-M_{11}+\Gamma_1^*(-M_{11})\Gamma_1)\ge\det(-M_{11}),\\
   \det(N_{22})=\det(M_{22}+\Gamma_2^*M_{22}\Gamma_2)\ge\det(M_{22}).
   \end{array}
\end{equation}
Hence
\begin{equation}
   \begin{array}{l}
   (-1)^m\det(M+\bar\Gamma
   M\Gamma)=\det(-N_{11})\det(N_{22}+N_{12}^*(-N_{11})^{-1} N_{12})\\
   \ge|\det M_{11}|\cdot|\det M_{22}|.
   \end{array}
\end{equation}
On the other hand,
\begin{equation}
   \begin{array}{l}
   (-1)^m\det M=\det(-M_{11})\det(M_{22}+M_{12}^*(-M_{11})^{-1} M_{12})\\
   \ge|\det M_{11}|\cdot|\det M_{22}|>0,
   \end{array}
\end{equation}
hence
\begin{equation}
   \det T\ge|\det V|^2\frac{|\det M_{11}|\cdot|\det M_{22}|}{|\det M|}.
\end{equation}
The lemma is proved.
\end{demo}

\begin{lemma}\label{lemma:asymp0}
Let $\Lambda=\diag(\Lambda_1,\Lambda_2)$ where $\Lambda_1$ and
$\Lambda_2$ are $n\times n$ diagonal matrices satisfying
$\bar\Lambda_2=\I\varepsilon\Lambda_1$ where $\varepsilon=\pm 1$.
Let $\zeta$ be a $2n$ dimensional column vector. For $j\ge k$,
denote
\begin{equation}
   R^\Lambda_{j\cdots k}=\left(\begin{array}{ccccc}\Lambda^j\zeta&\Lambda^{j-1}\zeta,
   \cdots,\Lambda^{k}\zeta\end{array}\right).
\end{equation}
Let
\begin{equation}\fl
   \begin{array}{l}
   \D\Pi^\Lambda=\left(\begin{array}{ccccccccccccccccc}
   \Lambda^n\zeta &\Lambda^{n-1}\zeta &\Lambda^{n-2}\zeta &R^\Lambda_{n-3\cdots 0}
   &K_n\bar R^\Lambda_{n-1\cdots 0} &0 &0\\
   \Lambda^{n+1}\zeta &\Lambda^{n}\zeta &\Lambda^{n-1}\zeta
   &0 &0 &R^\Lambda_{n-2\cdots 0} &K_n\bar R^\Lambda_{n-1\cdots 0}\\
   \end{array}\right),
   \end{array}\label{eq:Delta2}
\end{equation}
where $\D K_n=\left(\begin{array}{cc}&-I_n\\I_n\end{array}\right)$,
then $\det\Pi^\Lambda$ is purely imaginary.
\end{lemma}

\begin{demo}
$\bar\Lambda_2=\I\varepsilon\Lambda_1$ is equivalent to $\Lambda
K_n=-\I\varepsilon K_n\bar\Lambda$. Denote
$d^\Lambda=\det\Delta^\Lambda$. Multiplying row $j$ of $d^\Lambda$
by $-\lambda_j$ and adding it to row $2n+j$ $(j=1,\cdots,2n)$, we
get
\begin{equation}\fl
   \begin{array}{l}
   \D d^\Lambda=\left|\begin{array}{ccccccccccccccccc}
   \Lambda^n\zeta &\Lambda^{n-1}\zeta &\Lambda^{n-2}\zeta &R^\Lambda_{n-3\cdots 0}
   &K_n\bar R^\Lambda_{n-1\cdots 0} &0 &0\\
   0 &0 &0 &-R^\Lambda_{n-2\cdots 1} &\I\varepsilon K_n\bar R^\Lambda_{n\cdots 0}
   &R^\Lambda_{n-2\cdots 0} &K_n\bar R^\Lambda_{n-1\cdots 0}\\
   \end{array}\right|
   \end{array}
\end{equation}
by using $\Lambda K_n=-\I\varepsilon K_n\bar\Lambda$. Adding columns
$2n+2,\cdots,3n-1$ to columns $4,\cdots,n+1$, multiplying columns
$3n+1,\cdots,4n-1$ by $-\I\varepsilon$ and adding them to columns
$n+3,\cdots,2n+1$, we get
\begin{equation}
   \begin{array}{l}
   \D d^\Lambda=\left|\begin{array}{ccccccccccccccccc}
   R^\Lambda_{n\cdots 0}
   &K_n\bar\Lambda^{n-1}\bar \zeta &K_n\bar R^\Lambda_{n-2\cdots 0} &0 &0\\
   0 &\I\varepsilon K_n\bar\Lambda^{n}\bar\zeta &0 &R^\Lambda_{n-2\cdots 0}
   &K_n\bar R^\Lambda_{n-1\cdots 0}\\
   \end{array}\right|,
   \end{array}
\end{equation}
By moving the columns,
\begin{equation}
   \begin{array}{l}
   \D d^\Lambda=\left|\begin{array}{ccccccccccccccccc}
   R^\Lambda_{n\cdots 0} &K_n\bar R^\Lambda_{n-2\cdots 0} &0 &K_n\bar\Lambda^{n-1}\bar \zeta &0\\
   0 &0 &R^\Lambda_{n-2\cdots 0} &\I\varepsilon K_n\bar\Lambda^{n}\bar\zeta
   &K_n\bar R^\Lambda_{n-1\cdots 0}\\
   \end{array}\right|\\
   \D=\I\varepsilon\left|\begin{array}{cccccccc}
   R^\Lambda_{n\cdots 0} &K_n\bar R^\Lambda_{n-2\cdots 0}\end{array}\right|\cdot
   \left|\begin{array}{cc} R^\Lambda_{n-2\cdots 0} &K_n\bar R^\Lambda_{n\cdots 0}
   \end{array}\right|,
   \end{array}
\end{equation}
which is purely imaginary according to Lemma~\ref{lemma:conj}. The
lemma is proved.
\end{demo}

\section*{Acknowledgements}
This work was supported by National Basic Research Program of China
(2007CB814800) and STCSM (06JC14005).

\clearpage
\end{document}